\documentclass[bibyear]{aa}
\usepackage{graphicx}
\usepackage{txfonts}
\usepackage{amssymb}
\usepackage{lscape}

\newcommand{\kms}{km\,s$^{-1}$}

\newcommand{\HI}{H{\sc i}}

\newcounter{qub}

\begin{document}

\title{Study of galaxies in the Lynx-Cancer void}

\subtitle{VI. \HI-observations with the Nan\c {c}ay Radio Telescope}

\author{S.A. Pustilnik\inst{1,2}
          \and J.-M. Martin\inst{3,4} }

\institute{GEPI, Observatoire de Paris, Place Jules Janssen, 92195 Meudon,
                                                               France \\
   \email{sap@sao.ru}
\and
 Special Astrophysical Observatory of RAS, Nizhnij Arkhyz,
 Karachai-Circassia 369167, Russia
\and
  GEPI, Observatoire de Paris, PSL Research University, CNRS,
  Univ Paris Diderot, Sorbonne Paris Cit\'e, Place Jules Janssen, 92195
  Meudon, France
\and
Station de Radioastronomie de Nan\c {c}ay, Observatoire de Paris, CNRS,
F-18330 Nan\c {c}ay, France \\
   \email{jean-michel.martin@obspm.fr}
}

   \date{Received January 19, 2016; accepted July 26, 2016}


  \abstract
   {Void population consists mainly of late-type and low surface brightness
   (LSB) dwarf galaxies whose atomic hydrogen is the main component of
   their baryonic matter. Therefore, observations of void galaxy \HI\ are
   mandatory in order to understand  their evolution and dynamics.  }
   {Our aim was to obtain integrated \HI\ parameters for a fainter part of
  the nearby Lynx-Cancer void galaxy sample (total of 45 objects) with the
  Nan\c {c}ay Radio Telescope (NRT) and to conduct the comparative analysis
  of all the
   103 void galaxies with known \HI\ data with a sample of similar
  galaxies residing in denser environments of the Local Volume. }
   {For \HI\ observations we used the NRT with its sensitive antenna/receiver
   system FORT and standard processing. The comparison of the void and
   control samples on the parameter $M$(HI)/$L_{\rm B}$ is conducted with
   the non-parametric method , ``The 2$\times$2 Contingency Table test''.}
   {We obtained new \HI\ data for about 40~\% of the Lynx-Cancer galaxy
    sample. Along with data from the literature, we use these new data for
   further
   analysis of 103 void objects. The proxy of the evolutional parameter
   $M$(HI)/$L_{\rm B}$ of the void sample is compared with that of 82
   galaxies of morphological types 8--10 residing in the Local Volume (LV)
   groups and aggregates.}
   {At the confidence level of $P =$ 0.988, we conclude that for the same
   luminosity, these void galaxies are systematically gas-richer, on average
   by $\sim$39\%.
   This result is consistent with the authors' earlier conclusion on
   the smaller gas metallicities and evidence for the
   slower low-mass galaxy evolution in voids.}


   \maketitle

\section{Introduction}
\label{sec:intro}

Low-mass galaxies are thought to be the most fragile with respect to
both internal and external perturbations of various origin
(interactions, inflows, mergers) (e.g. Dekel \& Silk \cite{DS86},
Babul \& Rees \cite{Babul92}).
Therefore it is expected that their evolution is most sensitive to
various kinds of galaxy collisions (e.g. distant tidals, close pass-bys,
major and minor mergers). Thus, it may significantly depend
on the mean galaxy number density of various classes of the large-scale
structure.
Indeed, the strongest effects of dense environment on low-mass galaxy
properties are found in galaxy clusters (e.g. Boselli et al.
\cite{Boselli14} and references therein).
In this scheme, if external perturbations play a substantial role in the
secular evolution of typical dwarf galaxies, it is expected that
at least the
part of the dwarf galaxies located in voids could be less evolved objects.

In addition, studies of gravitational instabilities in the CDM
cosmology indicate that low-mass haloes become bound later in the
underdensity regions (low gravitational potential; e.g. Einasto et al.
\cite{Einasto11}). This is connected with the appearance of a bias in the
Gaussian peaks formalism for the structure formation (Bardeen et al.
\cite{Bardeen86}, Dekel \& Silk \cite{DS86}).
This, in turn, could also favour the appearance of less evolved low-mass
galaxies in voids.

The study of galaxies in voids was quite popular during the last decade
(Rojas et al. \cite{rojas05}, Patiri et al. \cite{Patiri06}, Sorrentino
et al. \cite{Sorrentino06}, Kreckel et al. \cite{Kreckel12}, Beygu et al.
\cite{Beygu12}, among others), thanks to the emerging large sky surveys like
the Sloan Digital Sky Survey (SDSS)  and the  Two-degree Field Galaxy
Redshift Survey (2dFGRS). However, most of the studies of void galaxies mentioned
above are mainly devoted to large and rather distant ($D \gtrsim$ 80--100
Mpc) voids.
This last choice, coupled with the demand of statistically complete galaxy
samples (based on the apparent magnitude limit),  limits the deepness of
their void galaxy samples at the level of $M_{\rm B} (M_{\rm r})
\lesssim$ --17.

Some differences between  void and wall galaxies in this luminosity range have
been found. In particular, void objects show a higher proportion of blue
galaxies and higher star formation rates (SFR)  (e.g. Rojas et al.
\cite{rojas04,rojas05}, and Hoyle et al. \cite{hoyle05,hoyle12}).
Similar results were obtained in the recent cosmological simulations by
Kreckel, Joung and Cen (\cite{Kreckel11}),
which indicate that the effect of the global environment of voids and walls
is rather
subtle for more massive galaxies. They found  evidence, however, that less
massive objects in voids can show signs of evolutionary youth.

In order to address the issue of void environment effect on low-mass galaxy
evolution (for galaxies below the adequate mass/luminosity limit,  e.g.
at $M_{\rm B} \sim$ --12 or fainter), it is necessary to study samples of
intrinsically faint objects.
 Having in mind the common apparent magnitude limits of the main
wide-angle redshift surveys (equivalent to $B_{\rm tot} \sim$ 18.0--18.5),
this implies the necessity to
study objects closer than $\sim$20~Mpc and located in the nearby voids which
are adjacent to the Local Volume.

In Pustilnik \& Tepliakova (\cite{PaperI}, Paper I) we described the large
galaxy
sample in the nearby Lynx-Cancer void ($D_{\rm centre} \sim$18 Mpc) and
presented their main known parameters. 
One of the tasks was to measure
and analyse the evolutionary parameters of void galaxies: metallicity
(or gas-phase O/H) and gas mass-fraction. In the published version, there
are 79 galaxies with the absolute magnitudes $M_{\rm B}$ in the range
[--12,--18.4], with a median of --14.0, and with the substantial
incompleteness at the fainter luminosities. Roughly
half of the void sample galaxies are low surface brightness (LSB) galaxies, 
with extinction and inclination corrected central SB values of
$\mu_{\rm 0,B,i} \geq $ 23.0$^{m}/\Box{\arcsec}$.

In Papers II and VII (Pustilnik et al. \cite{PaperII},2016), we present
a study of O/H in 81 members of the Lynx-Cancer void sample. We compared
the data with the parameter O/H of similar galaxies in denser environments.
Void galaxies appear to have systematically lower O/H (by about $\sim$37\%
in average) for the same luminosities. Other studies show that $\sim$20\% of
void LSB Dwarf galaxies (Pustilnik et al. \cite{PaperIII}; Chengalur,
Pustilnik \cite{Chengalur13}, Perepelitsyna et al. \cite{PaperIV}) are very metal-deficient and/or extremely gas-rich, indicating that
void environment is  conducive to unevolved objects.

Most of the late-type dwarf galaxies located in the Lynx-Cancer and
presumably in other nearby voids are LSB objects (e.g. Perepelitsyna et al.
\cite{PaperIV}, hereafter Paper~IV). They
are known to have a
significant or dominant part of baryon mass in the form of cold neutral gas.
To study the properties of this very important component of void galaxies,
it is necessary to know their global \HI\ parameters and, first of all, their
\HI\ mass in order to derive their second evolutionary parameter, the gas
mass-fraction $f_{\rm gas}$.
Moreover, since in many void galaxies the neutral gas appears to be
the main baryonic component, it is crucial to know its physical properties 
in order to understand galaxy dynamics and star formation.
Unusual, very gas-rich, and metal-poor galaxies found in course of \HI\
surveys, are good candidates for detailed \HI\ mapping.
Some of studies of very metal-poor dwarfs are presented in the papers of
Chengalur et al. (\cite{Chengalur06}) and  Ekta et al.
(\cite{Ekta06,Ekta08,Ekta09}).

For half of the void galaxy sample (mainly for the brighter, more
massive ones), the global \HI\ parameters were known from various
published sources (mainly from Haynes et al. \cite{ALFALFA},  Springob
et al. \cite{Springob05}, Huchtmeier \& Richter \cite{HR89}, Swaters et al.
\cite{Swaters02}, and Begum et al. \cite{Begum08}). For the remaining void
galaxies, we needed to conduct our own \HI\ observations.
Thus, the general goal of this work was to perform the most complete study
of the void galaxy \HI-properties. In addition, having the first results of
such
a study, there was a hope to find new  unusual very gas-rich dwarfs among
the fainter part of the void objects.

Some  very interesting void low surface brightness dwarfs (LSBDs), namely
very low metallicity and/or very gas-rich ones, were presented in Pustilnik
et al. (\cite{J0926,PaperIII}), Chengalur \& Pustilnik (\cite{Chengalur13}),
and Chengalur et al. (\cite{Chengalur15}).
As the data in Paper~I show, the Lynx-Cancer void galaxies have a rather
small radial velocity dispersion. This is interesting by itself in order to
make a comparison with cosmological simulations. This also relates directly
to the
identification of void filaments. Since \HI\ velocities are in general
substantially more accurate than optical ones, they provide an additional
opportunity to address the issues mentioned above.

Here we present all the NRT observed galaxies with known radial velocities
from the updated (relative to Paper~I) void sample (Pustilnik et al. 2016, in
preparation),  which currently includes  108 objects satisfying the
primary selection criteria of this sample.

\section{Sample}
\label{sec:sample}

In Table~\ref{t:Param} we present the main parameters taken from
NED\footnote{NASA/IPAC Extragalactic Database (NED)},
SDSS\footnote{Sloan Digital Sky Survey (Abazajian et al. \cite{DR7} and
references therein).},
or from the literature for all observed 45 void galaxies. New objects
taken from the updated Lynx-Cancer void sample (Pustilnik et al. 2016,
in preparation) are marked by an asterisk
(also in Table~\ref{t:Param_liter}). 
Table~\ref{t:Param} is organized as follows:
Col.~1 -- short IAU-style name;
Col.~2 -- other name or prefix (SDSS, HIPASS, etc.); Col.~3 -- galaxy type;
Cols.~4 and 5 -- Epoch J2000 R.A. and Dec; Col.~6 -- heliocentric velocity
from optical data (when available); Col.~7 -- heliocentric velocity from
\HI\ data; Col.~8 -- total $B$-band magnitude.
In most  cases, this value is calculated from the total $g$ ad $r$
magnitudes following Lupton et al. (\cite{Lupton05}).
These latter values
are obtained in Paper~IV on the photometry of the SDSS DR7
(Abazajian et al. \cite{DR7}) images.  For galaxies located outside of the
SDSS footprint, the $B$-band magnitudes are
adopted from Pustilnik \& Tepliakova (\cite{PaperI}) where respective
references are given. The only exception is J0802+0525 for which its
$B$-magnitude is estimated directly from its SDSS model $g$ and $r$ values
since we were unable to perform own photometry owing to a nearby bright star;
Col.~9 -- respective absolute magnitude, corrected for the Galaxy
extinction $A_{\rm B}$ according to Schlafly \& Finkbeiner (\cite{SF11}).
The adopted distances are based on heliocentric velocities $V$(\HI) from
Table~2. They are calculated according to the prescriptions given in Paper~I,
accounting for the large peculiar velocity in this region
$\Delta V \sim -300$~\kms\ (Tully et al. \cite{Tully08}).
For a few objects, distances were determined using  the velocity-independent
methods
(mainly Tip of RGB);  Col.~10 --  alternative name for the galaxy.

\section{Observations and reduction}
\label{sec:obs}

The \HI\ observations were made during the period 2007--2013 with the
Nan\c {c}ay  radio telescope (NRT). The NRT has a collecting area of
200$\times$34.5~m
and  a half-power beam width (HPBW) of 3.7$^{\prime}$~(east-west) $\times$
22$^{\prime}$~(north-south) at 21-cm and for a declination of
$\delta$=0$^\circ$ (see \verb|http://nrt.obspm.fr|).
A cooled 1.1-1.8 GHz dual-polarization receiver and a  8192-channel 
autocorrelation spectrometer were used for the observation of the HI line.
The system temperature was about 35~K
and the conversion factor of the antenna temperature to the flux density 
for a point source was 1.5~K~Jy$^{-1}$ near the equator.
The spectrometer covered a velocity range of about 2700~\kms, 
providing a channel spacing of 1.3~\kms\ before smoothing.
The effective resolution after averaging  four adjacent channels and
Hanning smoothing was $\approx$10.4~\kms.
Observations were obtained in separate cycles of `ON-source' and `OFF-source'
integrations, each of 40 or 60 seconds in duration. OFF integrations were
acquired at the target declination, with R.A. offset by
$\sim$15\arcmin~$\times cos(\delta)$ to the east.

A  noise diode was used to perform flux density calibration. Its power
was regularly monitored through the observations of known continuum and line
sources.
The comparisons of our measured fluxes with independent measurements of
the same objects with other telescopes indicates that flux density scales
are consistent within 10\%.

With the rms
noise of $\sim$1.5 to 5 mJy per resolution element after
smoothing (10.4~\kms), we achieved a S/N ratio for the peak flux densities
$F_{\rm peak}$ of the detected galaxies of 20--30 for the brightest objects,
while for the faintest sources we had detections with a S/N ratio of only
$\sim$2.5--4.
Total integration times per galaxy (ON+OFF+pointing time) varied between
0.6 and 6 hours.
For 3 of the 45 observed Lynx-Cancer void galaxies
we only obtained  upper limits of their $F_{\rm peak}$ and of their \HI\ flux.

Primary data reduction was made with the standard NRT program
{\it NAPS} written by the telescope staff.
The follow-up data processing was done with the IRAM package {\it CLASS}.
Both horizontal and vertical polarization spectra were calibrated and
processed independently. They were finally averaged together.
The baselines were generally well fitted by a third-order or lower
polynomial  and were subtracted out. Comments on the noise estimates and
on several marginally detected or undetected void galaxies of the 45 observed
are given in the next section.

\section{Results}
\label{sec:results}

Table~\ref{t:HI} presents the \HI\ parameters derived from the observations.
This is organized as follows:
Col.~1 -- short IAU-style name;
Col.~2 -- heliocentric velocity of the detected \HI\ line with its 1~$\sigma$
error, in \kms. This is determined as the midpoint between the half-peak
points on both sides of the \HI\ profile; Col.~3 -- the adopted distance as
in Paper~I (see comment for Col.~9 in Table~1); Cols.~4 and~5 -- velocity
widths
in ~\kms\ of the \HI\ profile at 50\% and 20\% of peak, $W_{\mathrm 50}$,
$W_{\mathrm 20}$ with their 1~$\sigma$ errors. They are determined as the
velocity range between the respective points on both sides of the \HI\
profile;
Col.~6 - $F$(\HI) -- integrated flux of detected \HI\ signal with its
1~$\sigma$
error in Jy~\kms. Formulae for error estimates of parameters in Cols.~2, 4--6
were adopted from our earlier NRT \HI-survey (Thuan et al. \cite{Thuan99}),
which
in turn uses the prescriptions from Schneider et al. (\cite{Schneider90});
Col.~7 -- logarithm of total mass $M$(\HI) in units of solar
mass with its 1~$\sigma$ error; Col.~8 -- ratio of $M$(\HI)/$L_{\rm B}$ with
related 1~$\sigma$ error, in solar
units; Col.~9 -- total time ON-source in minutes; Col.~10 -- rms of noise near
the \HI\ peak in mJy at the velocity resolution of 10.4~\kms; Col.~11 --
signal-to-noise ratio for peak value of the respective \HI\ profile.

\begin{figure*}
   \centering
 \includegraphics[angle=-0,width=18cm, clip=]{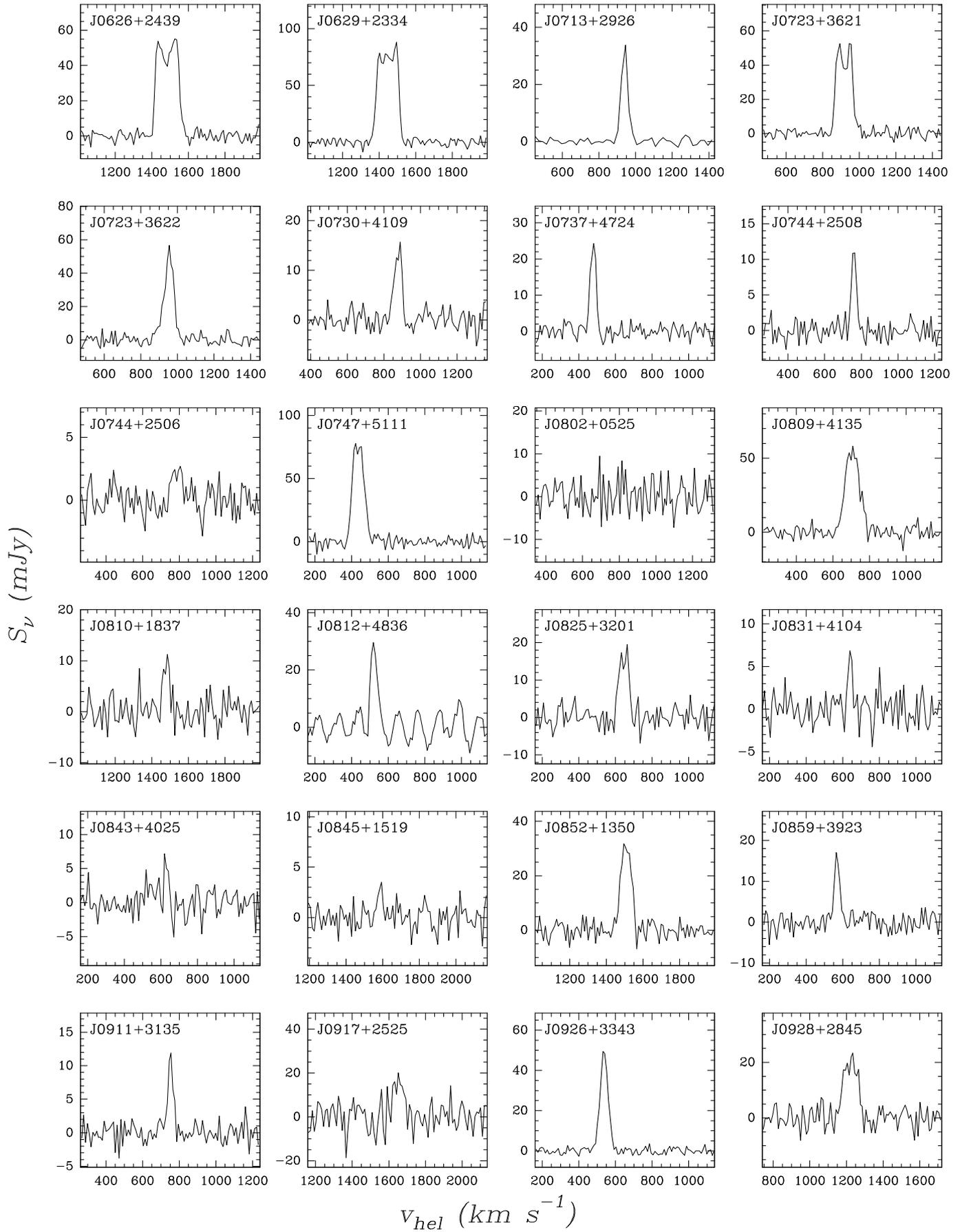}
   \caption{ NRT \HI\ profiles $S_{\nu}$ (in mJy) vs $V_{\rm hel}$
   (\kms) of all studied galaxies.
    }
         \label{Profiles1}
 \end{figure*}

\begin{figure*}
   \centering
 \includegraphics[angle=-0,width=18cm, clip=]{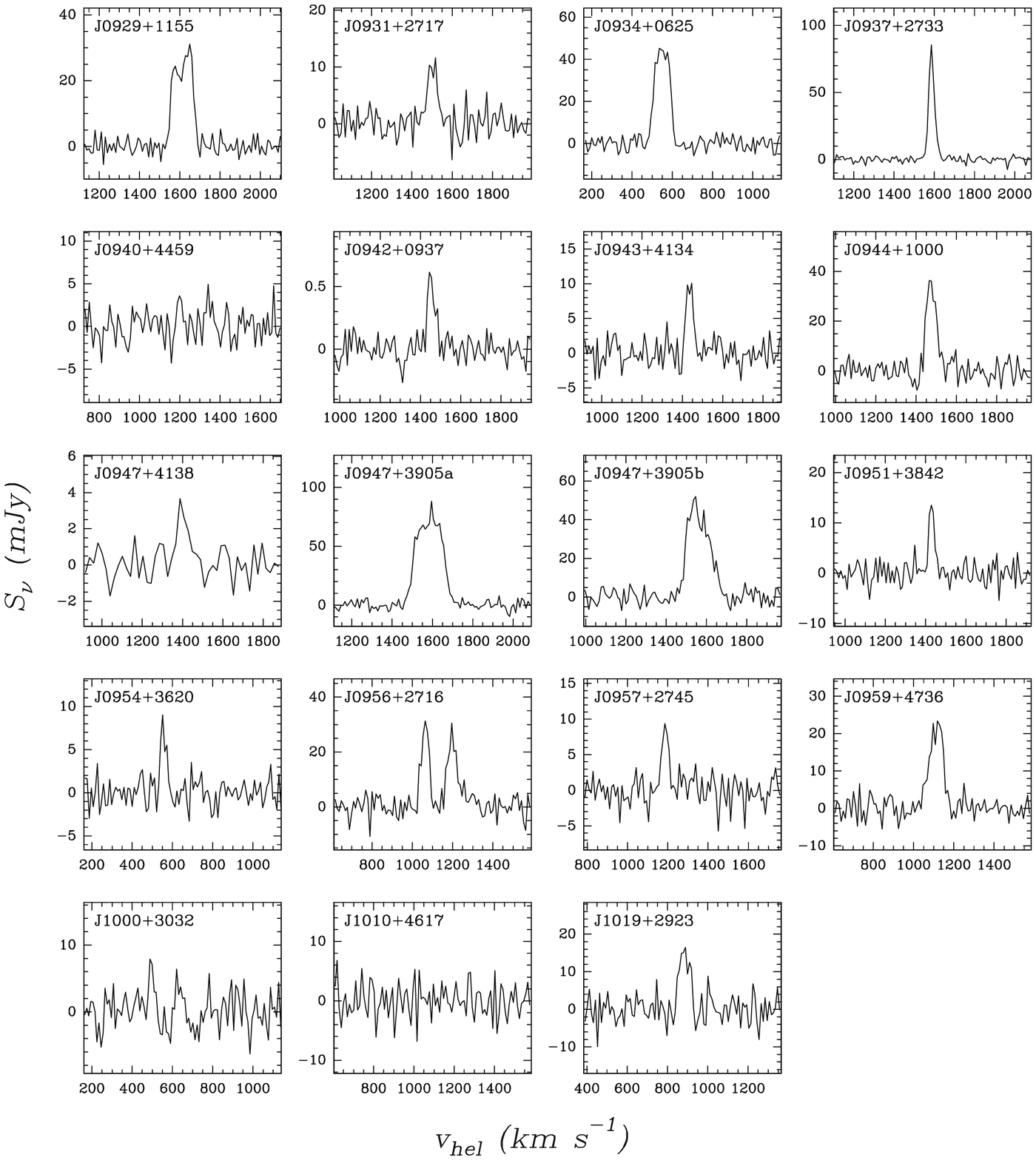}
   \caption{ NRT \HI\ profiles $S_{\nu}$ (in mJy) vs $V_{\rm hel}$
   (\kms) of all studied galaxies.
   }
         \label{Profiles2}
 \end{figure*}

Figures~\ref{Profiles1} and \ref{Profiles2} show in order of increasing R.A.
the \HI\ profiles of the void galaxies listed respectively in
Table~\ref{t:Param} and Table~\ref{t:HI}. For the triplet of
MRK~407 (J094747.60+390503.0) only two \HI\ profiles are presented;
they are rather complex and were obtained with the NRT beam pointing
in the direction of the largest members of the triplet. In addition, \HI\
profile of the galaxy J090018.30+322226.2 is not shown since no
signal was detected. Therefore, the  total number of profiles
displayed is 43.
We present below our comments about some peculiar objects.

{\it SDSS~J072301.42+362117.1} and {\it J072313.46+362213.0}. The first
galaxy, a LSB dwarf, has been identified as a new Lynx-Cancer void galaxy
after its redshift was obtained at the SAO 6m telescope (BTA).
Its NRT \HI\ profile suggested a
possible contribution from a nearby galaxy which has been found as a very
low surface brightness dwarf, $\sim$2$^m$ fainter than the main component,
at $\sim$3\arcmin\ to E (J072313.46+362213.0).
An additional NRT observation in the direction  partly disentangled 
the confusion between the \HI\ contributions. The final \HI\ parameters for
these two galaxies and a third  even fainter one (J072320.57+362440.8,
see Table 3) have been adopted after the subsequent GMRT \HI\ mapping of this
triplet  (Chengalur, Pustilnik \cite{Chengalur13}).

{\it SDSS J080238.15+052551.2}. This very faint and compact optical object
close to a bright star was included in the void sample after its assumed
identification with a faint ALFALFA source (Haynes et al. \cite{ALFALFA}) with
$V$(\HI)=830~\kms, $F$(\HI)=0.43$\pm$0.04~Jy~\kms\ with S/N ratio of 6.2
(AGC~188988). With such parameters, this galaxy has a very high ratio
$M$(\HI)/$L_{\rm B}$ $\sim$4.8 (using the total SDSS $g$ and $r$ magnitudes,
transformed to $B_{\rm tot}$=19.8).
Since very gas-rich galaxies are rare objects, we conducted \HI\ 
observations of AGC~188988 with the NRT.
Our NRT data indicate no signal at the respective velocity with an upper
limit of $F$(\HI) $<$0.16~Jy~\kms\ (2$\sigma$).

The probable interpretation of this case is a false ALFALFA detection.
If, however, there is \HI\
associated with the suggested faint optical object, its ratio
$M$(\HI)/$L_{\rm B} <$1.8  is  not so large. We consider this object's data
to be unreliable and excluded it from the following statistical analysis.

{\it SDSS J090018.30+322226.2}. This galaxy is a new void object with
$V_{\rm hel}$ = 740$\pm$30~\kms, as measured on the faint H$\alpha$ emission
in the spectrum obtained at the BTA. At NRT the signal at this radial velocity
is within 1.4$\sigma$, so the numbers below should be treated as upper
limits. With $W_{\mathrm 50}$ = 30~\kms, typical of galaxies with
$M_{\rm B} \sim$ --12, which results in the total flux of
0.14$\pm$0.10~~Jy~\kms. For parameter  $M$(\HI)/$L_{\rm B}$
the respective value is 0.72$\pm$0.52.

{\it SDSS J094003.27+445931.7}. This galaxy has marginally detected \HI\ with
the radial velocity close to that derived from the SDSS emission-line spectrum
(1358$\pm$4~\kms). However, there is also a similar flux detection (at the
level of $\sim(2-3)\sigma$) at $V_{\rm hel} =$ 1202$\pm$8~\kms. The search for
a possible optical counterpart on the SDSS image to this \HI\ component
produced two candidates within the NRT beam.

The nearest one is a small and almost edge-on blue disc SDSS
J093951.28+445921.9 with $g$=19.34, $r$=19.15. Its BTA spectrum has revealed
the velocity of the H$\alpha$ line of $\sim$14000~\kms.
The second candidate is the almost face-on LSB disc SDSS J093950.11+444800.1
with $g$=17.55, $r$=17.18 ($B \sim$17.90), $\sim$11.5\arcmin\ to the south and
13.16~s ($\sim$140\arcsec) to the west. Owing to the NRT beam offset, its
nominal
\HI-flux should drop by a factor of 6.5. In this case, its ratio
$M$(\HI)/$L_{\rm B}$ is $\sim$1.4, rather typical of void LSBDs.
The optical redshift of this galaxy is needed in order to fix
the origin of the second \HI\ source.

{\it MRK~407=J094747.60+390503.0}. This blue compact galaxy (BCG) is the
brightest member of a triplet, which also includes  $\sim$1.7 mag fainter LSBD
UZC~J09475+3908 at 3\arcmin\ to the north and  $\sim$3 mag fainter LSBD
SDSS J094758.45+390510.1 at $\sim$2\arcmin\ to the east. Each galaxy
contributes
to the $F$(\HI) for any NRT pointing in the direction of the triplet.
We used our NRT results, accounting for the a priori known decrease of
\HI\ flux for the sources NRT beam offset, as well as the earlier
observations of MRK~407 by Thuan \& Martin (\cite{TM81}) to disentangle the
contribution of each component of the triplet. The typical estimated accuracy
of the resulted $F$(\HI) is
$\sim$20\%. Follow-up GMRT mapping of this triplet (Chengalur et al.,
in preparation) will give a better understanding of its properties.

{\it SDSS~J095633.65+271659.3} with $V_{\rm hel} =$1059~\kms\  is a faint
companion of a $\sim$4$^m$ brighter spiral IC~2520 (at 13.2$^s$ to the west
and 3.3\arcmin\ to the south, see Table~3), which is also in the NRT beam
and is seen in the plot of the \HI\ profile as an additional peak
  at $V_{\rm hel} =$1243~\kms.

{\it SDSS J101014.96+461744.1}.  This is a faint galaxy with a good S/N SDSS
emission-line spectrum. Its optical redshift corresponds to
$V_{\rm hel} =$1092$\pm$3~\kms. On our data, there is no
detectable \HI\ flux, for $\sigma_{\rm noise} \sim$2.4~mJy.
For statistical analysis, we adopt for its $F$ the upper limit
$F$(\HI)$<$0.12~Jy~\kms\ and the value of  $M$(\HI)/$L_{\rm B} <$0.34.

\section{Analysis}
\label{sec:analysis}

In our analysis of the properties of the Lynx-Cancer void galaxy sample, 
we use the optical parameters gathered from the literature, as mentioned
 in section 2, and the \HI\ parameters obtained from our
observations or taken from the literature.
Table~3 lists galaxies with \HI\ data taken from the literature, with their
\HI\ and optical parameters. Table~3 is organized as follows:
Cols.~1 to 5 -- same as Table~\ref{t:Param};
Col.~6 -- adopted heliocentric velocity $V_{\rm hel}$;
Cols.~7 and 8 -- the total apparent and absolute $B$ magnitudes: for the
published sample outside the SDSS zone,   from Paper~I; for galaxies with
new photometry,  from the papers by Perepelitsyna et al. (\cite{PaperIV});
and for the rest of the updated version of the void sample,  from Pustilnik
et al. (2016, in preparation). The few exceptions are the following.
For J0706+3620 and UGC~3672, their $B$ magnitudes are adopted from Chengalur
et al. (\cite{CPE16}). For J0736+0959, its $B$ magnitude is adopted from the
recent photometry in Haurberg et al. (\cite{Haurberg15}).
For J0956+2900 (DDO68C), there is no possibility of estimating its optical
flux because of a nearby bright star.
We adopt its $B$ magnitude based on $M$(\HI)
in Cannon et al. (\cite{Cannon14}) and a typical of this sample value of
$M$(\HI)/$L_{\rm B}$=1;
Cols.~9 and 10 -- total \HI\ flux $F$(\HI) and derived $M$(\HI);
Col.~11 -- reference to the \HI\ data.
In total, we use the data of 103 void galaxies in this analysis.

In order to compare the gas-content parameter $M$(\HI)/$L_{\rm B}$ of the
void sample with that of a sample of galaxies in denser environments,
we created
a sample of 82 late-type dwarf and subluminous galaxies in the LV
residing in groups and the Canes Venatici I (CVnI) cloud.
The latter were described by Karachentsev  (\cite{Kara_groups05}).
We used those members of these groups for which we found \HI\ data in the
literature, mainly in the Catalog of Nearby Galaxies (CNG)
by Karachentsev et al. (\cite{CNG}).

\begin{figure*}
   \centering
 \includegraphics[angle=-90,width=8cm, clip=]{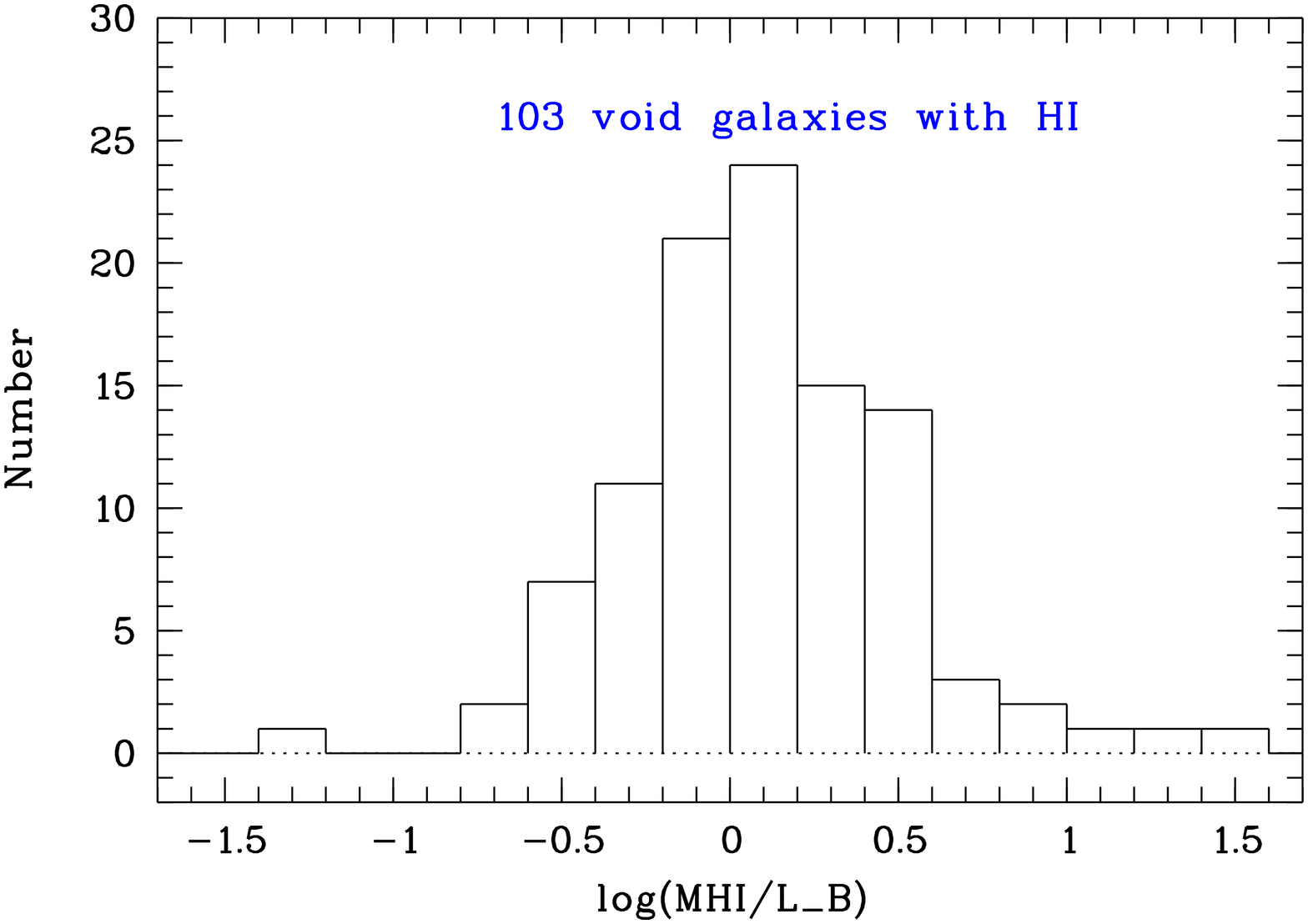}
 \includegraphics[angle=-90,width=8cm, clip=]{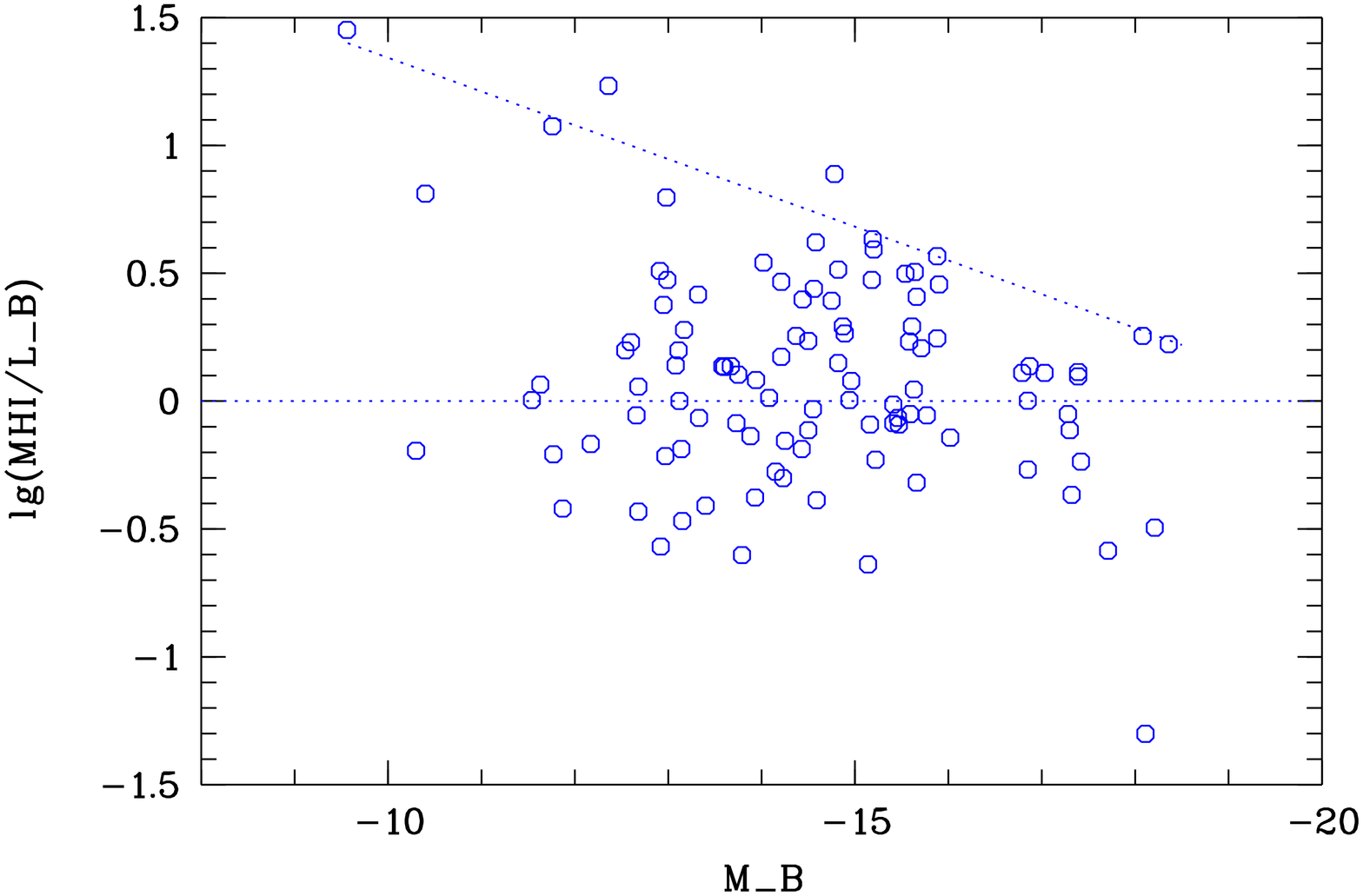}
 \includegraphics[angle=-90,width=8cm, clip=]{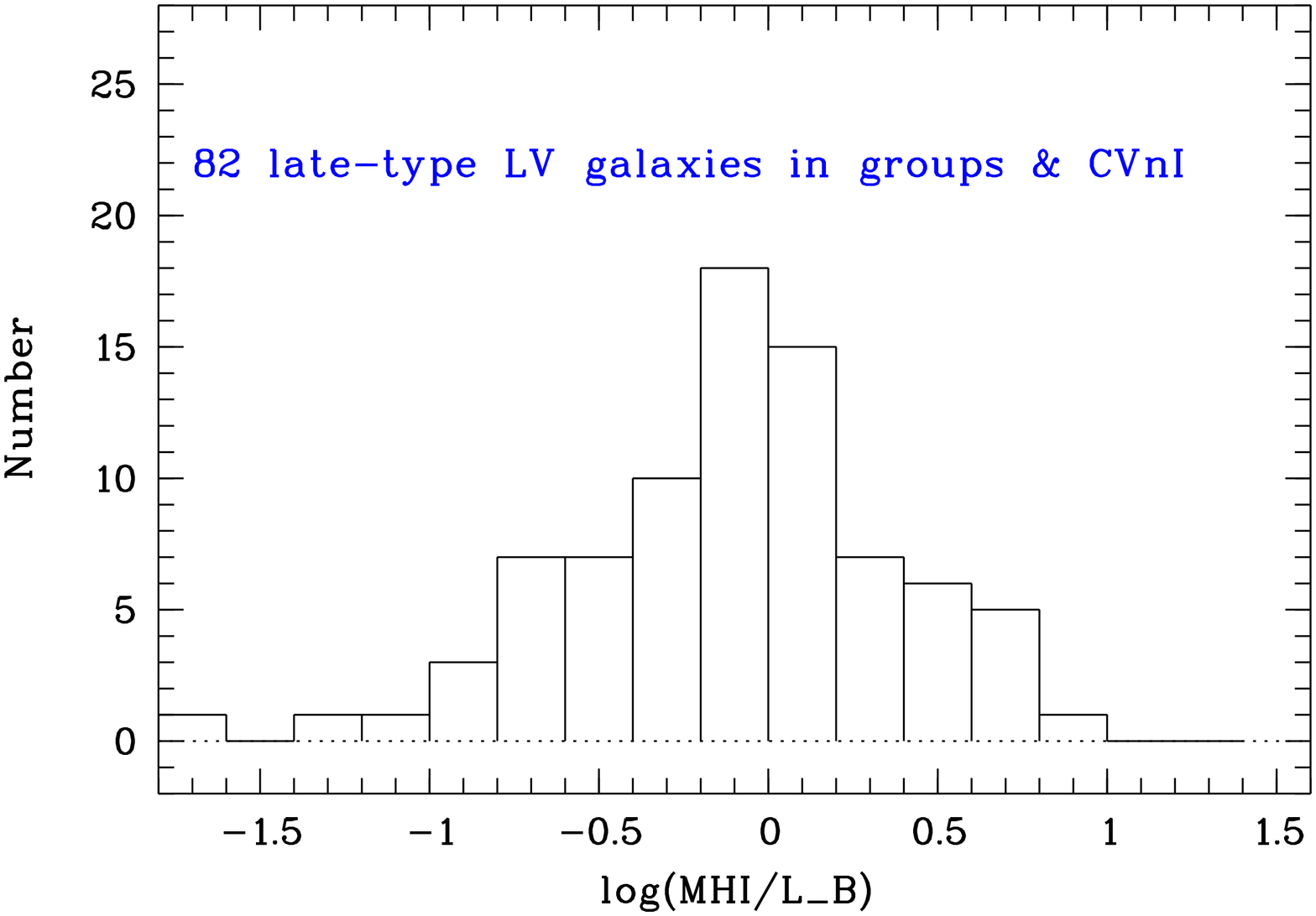}
 \includegraphics[angle=-90,width=8cm, clip=]{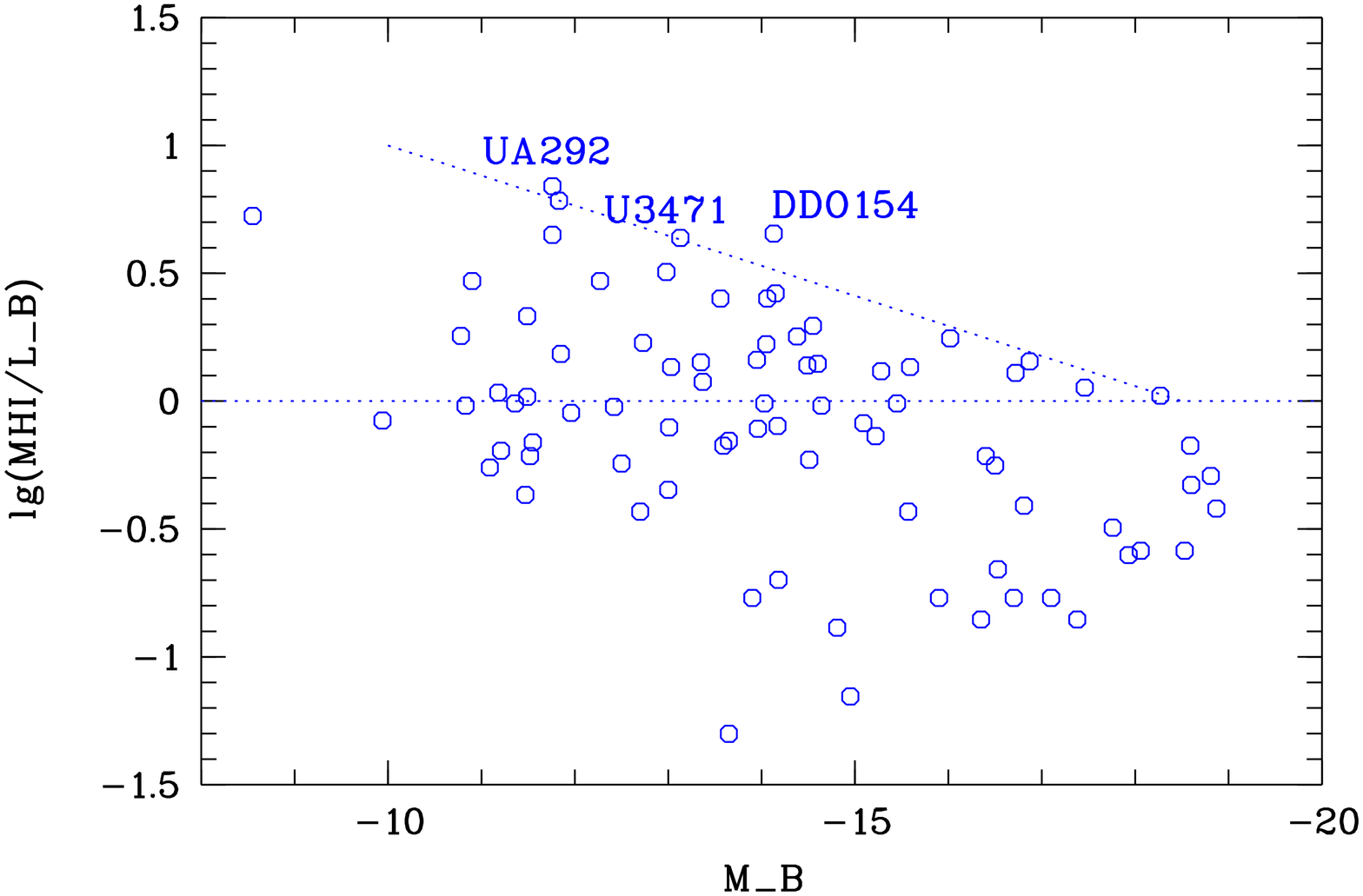}
   \caption{{\bf Top left:} Distribution of mass-to-light ratio
   $M$(HI)/$L_{\rm B}$ for all Lynx-Cancer void galaxies with \HI\ data
and {\bf Top right:}
    Relationship between $M$(HI)/$L_{\rm B}$ and the absolute magnitude
   $M_{\rm B}$.
   {\bf Bottom left:} Distribution of $M$(HI)/$L_{\rm B}$ for
   late-type dwarf and subluminous galaxies in the Local Volume (LV) groups
   and the CVnI cloud from  Karachentsev (\cite{Kara_groups05}).
   {\bf Bottom right:}
    Relationship between $M$(HI)/$L_{\rm B}$ and the absolute magnitude
   $M_{\rm B}$ for the same late-type galaxy sample. The scatter of
   parameter $M$(HI)/$L_{\rm B}$ is large for all values of $M_{\rm B}$,
   indicating the interplay of several significant factors. Nevertheless, the
   fraction of
   higher $M$(HI)/$L_{\rm B}$ ratio objects is clearly larger for void
   galaxies. Also, the three highest ratio galaxies from the LV sample are
   situated in the outer parts of groups and CVnI cloud (marked by their
   names).  Median values of
   $M$(HI)/$L_{\rm B}$ are respectively, 1.21 and 0.87 for void and the LV
   `late-type in groups' samples, which differ by a factor of $\sim$1.39.
   The upper boundary line for the void sample  is also a factor of 1.6--2
   higher than that  for the `group' sample.
}
         \label{histogram}
 \end{figure*}

We show the distributions of parameter $M$(\HI)/$L_{\rm B}$ for the void
and the  LV-groups samples in Figure~\ref{histogram} (top and  bottom
left panels, respectively).
The median value of $M$(\HI)/$L_{\rm B}$ for the combined void and
LV-group sample of 185 galaxies is equal to 1.01. Therefore, galaxies with
$M$(\HI)/$L_{\rm B} \geq$1.0 are denoted  `gas-rich'.
Although each sample shows rather large scatter
(indicating that there are several affecting factors), the distribution
of the void galaxies is confidently shifted to the higher values of
$M$(\HI)/$L_{\rm B}$.
This effect is apparent in a factor of $\sim$1.39
difference between their medians (1.21 and 0.87, respectively) and the
significant difference in fractions of gas-rich
objects in the void and LV-groups samples (0.59 and 0.41).
Since these differences might be due to the statistical scatter, more
advanced statistical tests are needed.

The significance of the second difference can be tested via non-parametric
statistical methods. In particular, we use a test well known in biology and
quality control, called The 2$\times$2 Contingency Table test\
(e.g. Bol'shev and Smirnov \cite{BoSm83}, and references therein).
It appears to be more powerful than the  Kolmogorov--Smirnov test in problems
like this, as was tested for a similar astronomical problem in Pustilnik
et al. (\cite{Pustilnik95}, see the detailed Appendix).
Here we briefly summarize the process of grouping the galaxies in the
respective cells of \mbox{Table 2$\times$2}.
The zero hypothesis $H_{\mathrm 0}$ states that the property $G$ being 
`gas-rich' does not relate to the
property $V$  belonging to the void environment, or in other words, the
fraction of gas-rich galaxies is the same for both compared samples of
late-type galaxies. The respective Table 2$\times$2 reads as follows:

\begin{center}
\begin{tabular}{llll} \hline \hline
Property   & $G$   & non-$G$ & Sum~1  \\   \hline
$V$        & $m$=61 & $n-m$=42 & $n$=103   \\
non-$V$    & $M-m$=34 & $N-n-(M-m)$=48 & $N-n$=82  \\   \hline
Sum~2     &  $M$=95  & $N-M$=90 &  $N$=185     \\
\end{tabular}
\end{center}

Here $m$=61 is the number of void gas-rich galaxies and $n-m$=42 is the number
of void non-gas-rich objects;  \mbox{$M-m$=34} is the number of non-void
gas-rich galaxies, while $N-n-(M-m)$=48 is the number of non-void non-gas-rich
ones. If properties $G$ and $V$ were independent, that is, if there were no
correlation between the property of belonging to void and the property of
being gas-rich, the probability of accidently getting a table with the same
occupation
numbers  [61, 42, 34, 48], calculated according to the formulae in the
above Appendix (and Bol'shev and Smirnov \cite{BoSm83}) is less than
$p=0.012$.
The respective probability of rejecting $H_{\mathrm 0}$ with the given
occupation numbers is $P = 1-p =$ 0.9882.
Hence, the visual impression on the significantly
higher fraction of void gas-rich objects is supported by the statistical
criterion at the confidence level $P$ of 0.9882.

In addition to the distributions on parameter $M$(\HI)/$L_{\rm B}$, 
in the  right-hand panels of Fig.~\ref{histogram} we also
compare its relation to galaxy
luminosity (via parameter $M_{\rm B}$). To aid the eye, we draw upper
boundary straight lines for both samples. The visual inspection shows that for
the Lynx-Cancer void galaxies this upper line goes slightly higher (by a
factor of 1.6--2.0) than for
`group' galaxies in the whole range of galaxy luminosities. The same is valid
for the bottom boundary. With only one exception, the most gas-poor void
galaxies have substantially higher values of $M$(\HI)/$L_{\rm B}$ than
the similar galaxies in groups.

\section{Discussion}
\label{sec:discussion}

In discussing noticeable differences in the gas content between void and
group late-type galaxies, it is important to pay attention that the
group  sample is itself quite inhomogeneous, including  the
classical Local Volume groups similar to our own Local Group (M81, CenA, M83,
IC~342, Maffei, Scu) one rather rarefied and unrelaxed aggregate known as the
CVnI cloud. It is curious and instructive that three of the
six most gas-rich galaxies in the group sample ($M$(HI)/$L_{\rm B}=4.3-6.9$)
belong to the outer parts of this aggregate, and hence
can be treated as falling to this from the lower-density environment.
These most gas-rich galaxies include DDO~154, UGCA~292, and UGC~3741 with
respective values of parameter $M$(HI)/$L_{\rm B}$ of 4.5, 6.9, and 4.3
(see the recent summary in Chengalur, Pustilnik \cite{Chengalur13}).
The structure of the CVnI cloud was revisited by Makarov, Makarova and
Uklein (\cite{Makarov13}), based on the improved TRGB distance determinations.

The above-mentioned  extreme members of the CVnI cloud reside far from the
centre of the cloud, closer to the zero-velocity radius ($R$=1.06~Mpc)
or substantially further (at $\sim$1.1, 0.9, and 1.6 Mpc, respectively),
and thus can probably be treated as being in the process of fall-off onto
CVnI cloud. Coming from the significantly lower density environment, they
can possess properties of some of the most unevolved representatives of
underdense regions.

To check the effect of the CVnI cloud galaxies on the comparison of the LV
vs Lynx-Cancer void galaxy samples, we have removed the 11 CVnI cloud
galaxies from the whole Local Volume sample and are left with
71 LV galaxies. We apply the same Table 2$\times$2 method used above
to check the null hypothesis $H_{\mathrm 0}$ on the independence of a gas-rich
galaxy fraction on the type of environment.
For the new table, the probability of  accidently getting the variant
with occupation numbers [61, 42, 27, 44] is $p$ = 0.00444,
$\sim$2.7 times smaller than for the whole Local Volume subsample.
The respective confidence level to reject $H_{\mathrm 0}$ is
$P = 1-p$ = 0.99556.
This probably indicates that at least a part of the CVnI cloud galaxies are in
a special evolutionary status.

\begin{figure*}
   \centering
 \includegraphics[angle=-90,width=8cm, clip=]{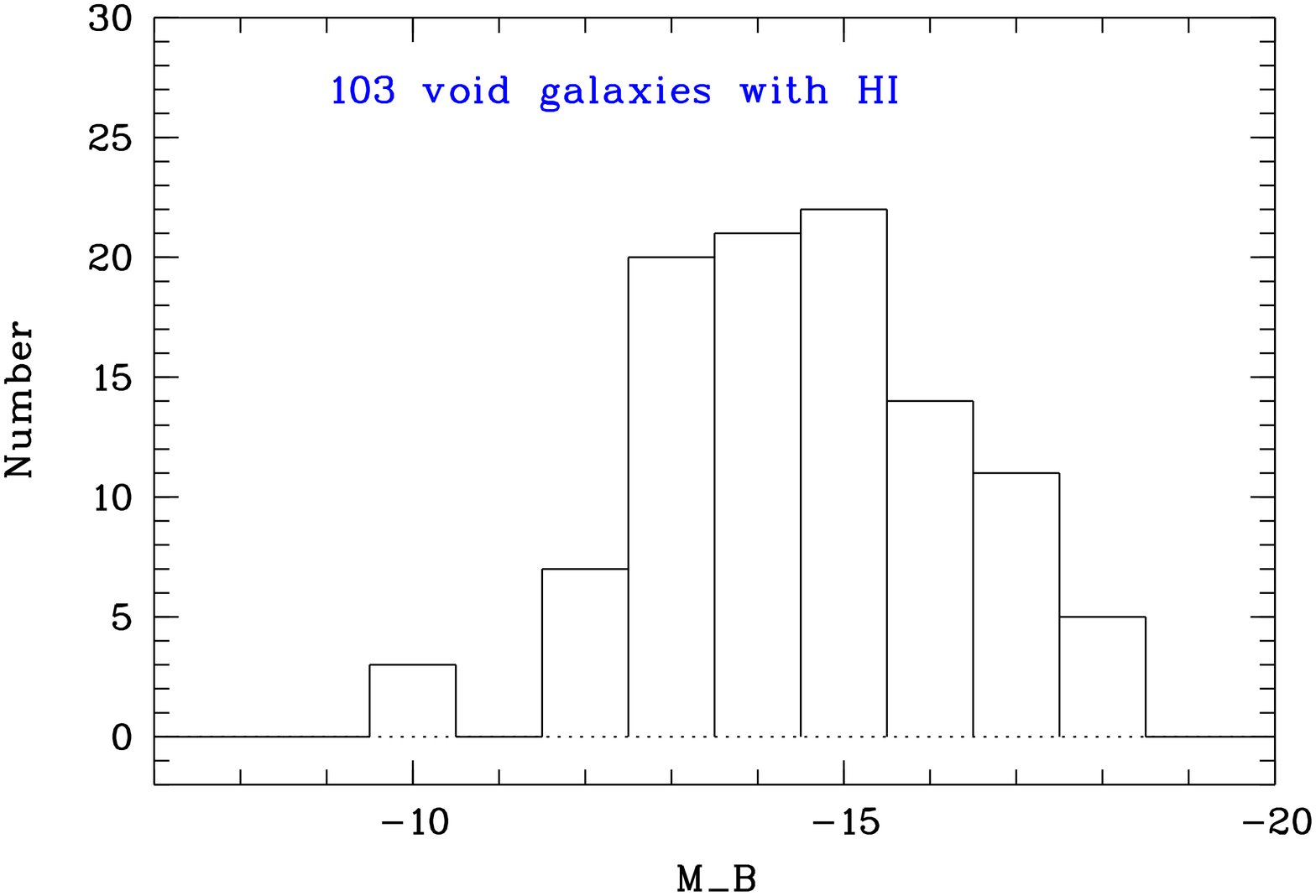}
 \includegraphics[angle=-90,width=8cm, clip=]{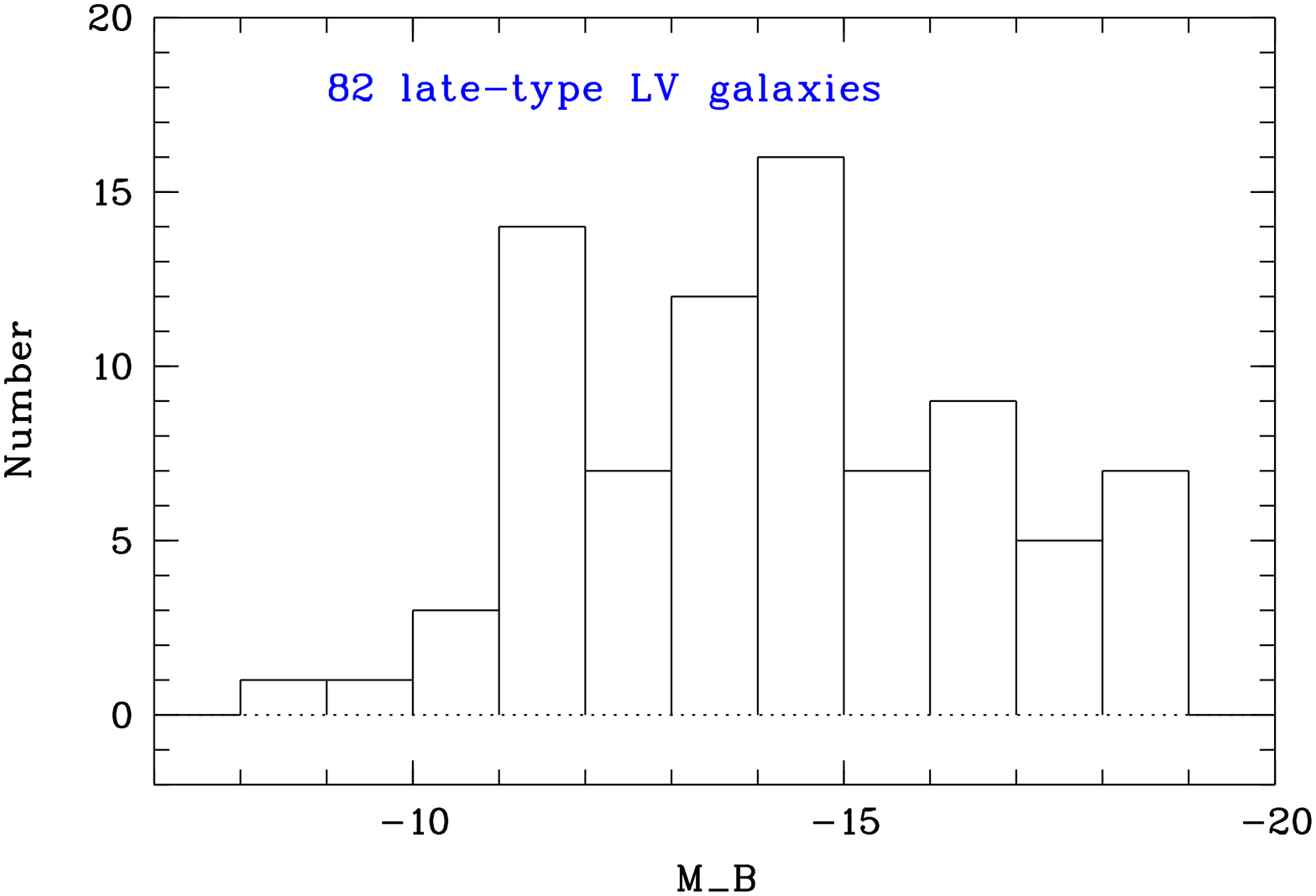}
   \caption{{\bf Left:}  Distribution of $M_{\rm B}$
   for 103 Lynx-Cancer void galaxies with \HI\ data.
 {\bf Right:}  Same distribution of $M_{\rm B}$ for the comparison sample
 82 galaxies in groups inside the Local Volume.
   Median and mean values of $M_{\rm B}$ of the groups sample (--14.10
   and --14.24) are somewhat fainter than for the void sample (--14.45 and
   --14.61, respectively). The standard deviation for the former, in contrast
is somewhat
   higher (2.38 mag vs 1.83 mag). See the Discussion for further implications.
}
         \label{histoMB}
 \end{figure*}

\begin{figure*}
   \centering
 \includegraphics[angle=-90,width=8cm, clip=]{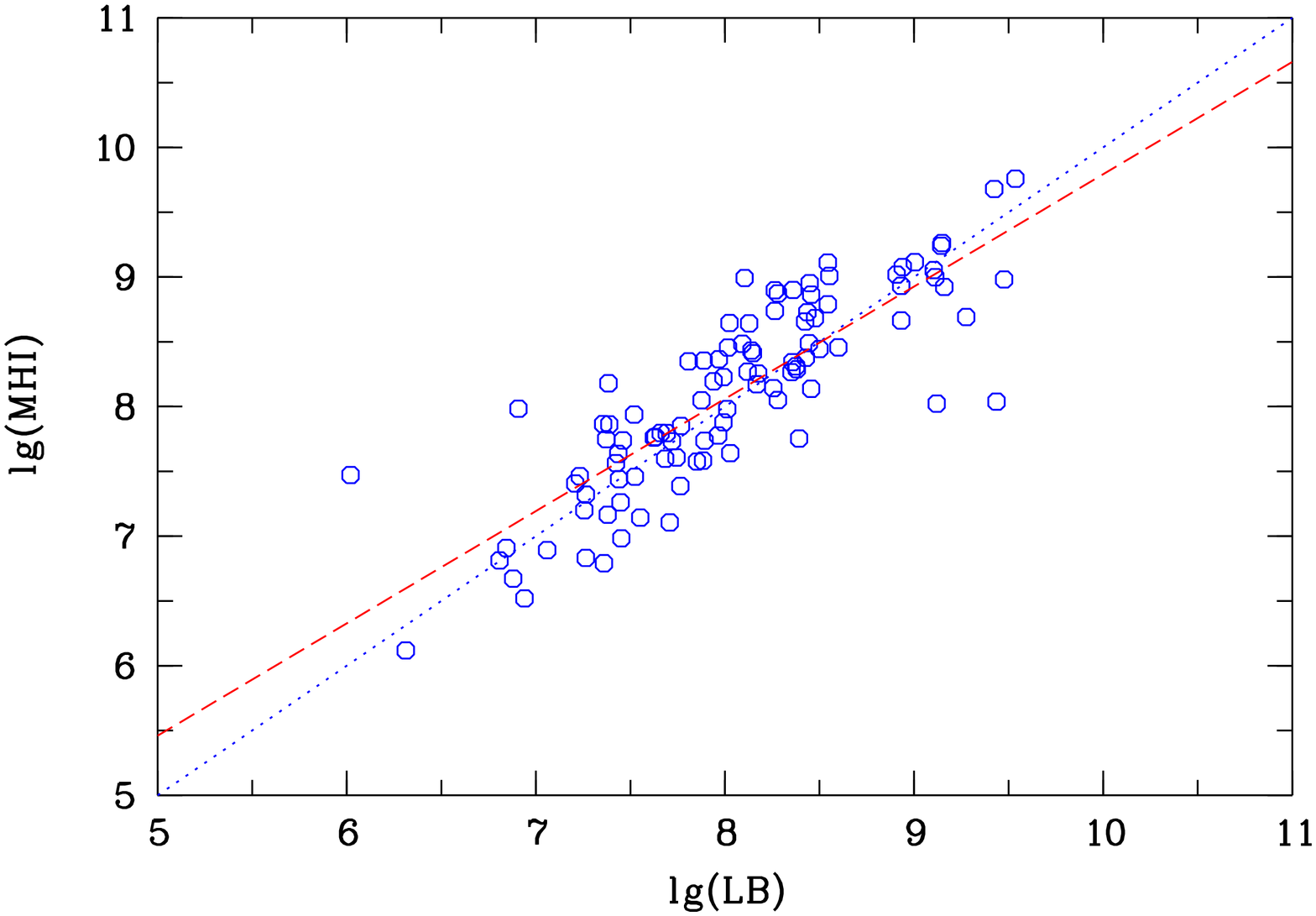}
 \includegraphics[angle=-90,width=8cm, clip=]{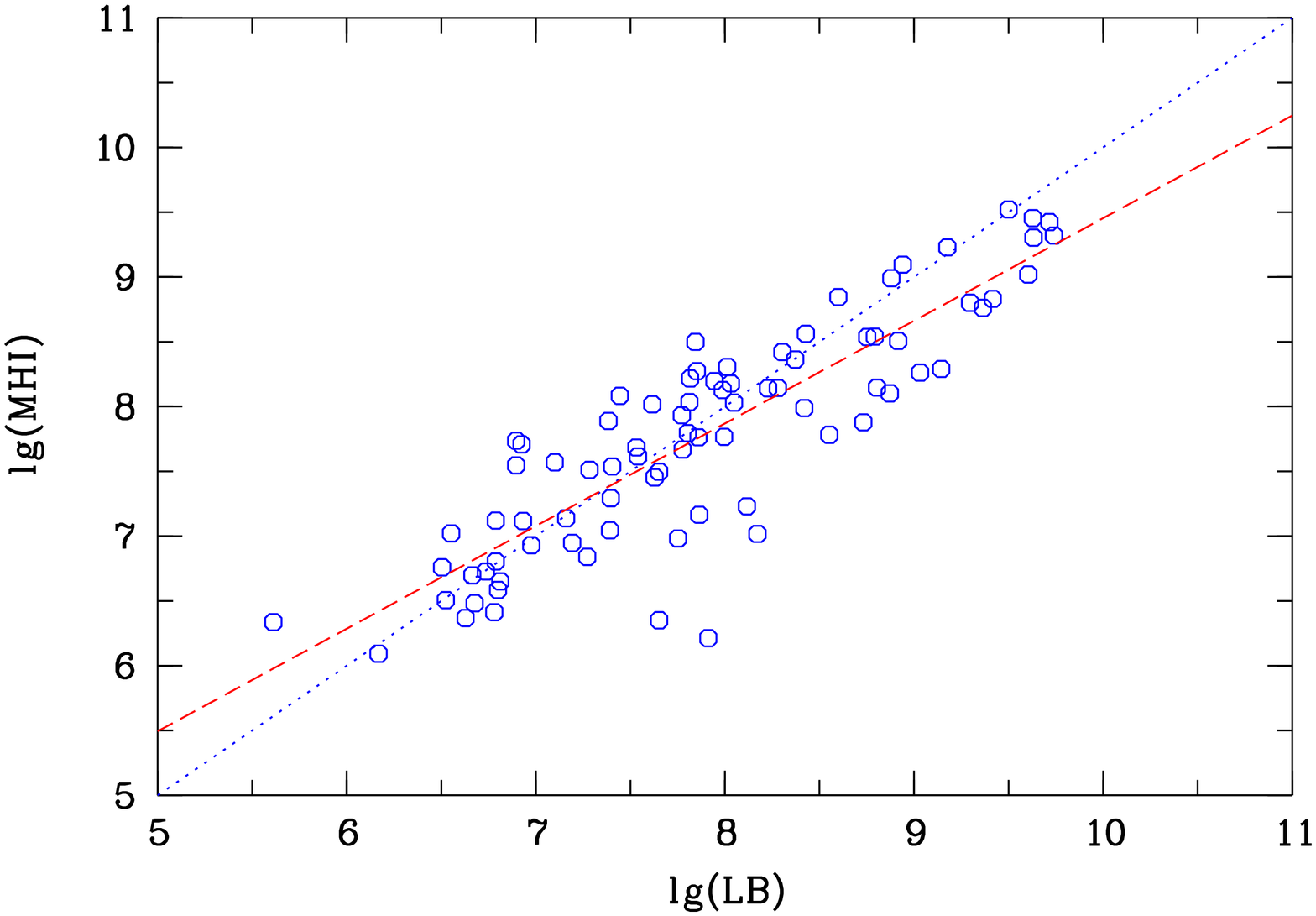}
 \caption{{\bf Left:} Relation between $M$(\HI) and $L_{\rm B}$ (in solar
  units) for 103 Lynx-Cancer void galaxies with \HI\ data. The dotted line
   shows
   positions for objects with $M$(\HI)/$L_{\rm B}$=1. The dashed line shows
   the linear regression on all galaxies, with the slope of
   $k1$=0.875$\pm$0.055 and rms=0.45 (in $\log$$M$(\HI)).
 {\bf Right:}  Same relation for all 82 galaxies from the comparison sample
  in the Local Volume groups. The slope of the linear regression is
  $k2$=0.792$\pm$0.051  and rms=0.44 (in $\log$$M$(\HI)).
  See the Discussion for further implications.
}
         \label{regression}
 \end{figure*}

One important note relates to the conclusion on the difference in
distributions of parameter $M$(HI)/$L_{\rm B}$ for the void and group
(Local Volume groups) samples. As  can be seen in the right panels
of Fig.~\ref{histogram},
there is a trend (also known  from several earlier works, see e.g. Huchtmeier
et al. \cite{Hucht97}; Pustilnik et al. \cite{Pustilnik02}): the ratio
$M$(HI)/$L_{\rm B}$ increases
with the decrease in galaxy luminosity (see also Fig.~\ref{regression} for the
rate of this increase). Therefore, if the two samples under comparison differ
significantly in $M_{\rm B}$ distribution, it is possible to obtain a
noticeable
difference in distribution $M$(HI)/$L_{\rm B}$, even though in reality these
samples have the same distribution. In
Figure~\ref{histoMB} and with related nubmers we show that this is not the
case.
Indeed, both distributions on $M_{\rm B}$ are  similar, have close mean
and median
values of $M_{\rm B}$ (see numbers in the Figure legend), and are somewhat
lower for the group sample. The latter should lead in general to the
opposite effect, that is the group sample should have more numerous gas-rich
galaxies.

In Fig.~\ref{regression} we show how the galaxy hydrogen mass $M$(\HI)
is related to the blue luminosity $L_{\rm B}$. The left panel is for
the sample of 103 Lynx-Cancer void galaxies, while the right panel is for
82 galaxies of the Local Volume group sample. Dotted lines show the positions
of galaxies with $M$(\HI)/$L_{\rm B}$=1 (in solar units, with a slope of 1.0).
The red dashed lines (see figure legend) show the real linear regression for
considered samples.
They indicate that for both samples, galaxies become on average gas-richer
with decreasing luminosity. The respective coefficients in the relation
log($M$(\HI)/$L_{\rm B}$)/log($L_{\rm B}$)=--0.129$\pm$0.054 (void sample)
and
--0.208$\pm$0.051 (Local Volume sample) do not differ significantly. Thus,
their average $<k>$= --0.163$\pm$0.040 can be considered  representative
of such a relationship for both samples.

It is interesting to compare this result with estimates published for other
samples. In particular, Staveley-Smith et al. (\cite{Staveley92}) found
$k$= --0.3$\pm$0.1
for a sample of LSB dIs and BCGs, while Smoker et al. (\cite{Smoker00}) found
$k$= --0.2$\pm$0.1 for emission-line galaxies of the University of Michigan
survey.  In the Pustilnik et al. (\cite{Pustilnik02}) study of BCGs in
various environments, this
slope for non-cluster BCGs is consistent with average of $<k> = -0.25\pm$0.1.
Thus, within rather large scatter, all available data for late-type and BCG
galaxies on the relation $M$(\HI)/$L_{\rm B} \varpropto$ $L_{\rm B}^{k}$
are consistent with the common index $k \sim$ --0.2. This corresponds to a
factor of $\sim$4 increase in $M$(\HI)/$L_{\rm B}$ for a
factor of 1000 decrease in luminosity.

This relation is a specific illustration of the well-known down-sizing
phenomenon which is connected to the slower evolution of smaller mass
galaxies. The general trend towards lower gas metallicity for smaller
galaxies is a better-known manifestation of the same phenomenon.

\subsection{Summary}
\label{sec:summ}

\begin{enumerate}
\item
 NRT \HI\ data are presented for 45 galaxies of the Lynx-Cancer void.
Along with \HI\ data already published in the literature, we were able to build a
large sample of 103 galaxies and study the properties of $\sim$95\% of the
updated Lynx-Cancer void galaxy sample.
\item
The analysis of parameter $M$(\HI)/$L_{\rm B}$ -- observational proxy of the
relative gas content -- for the void galaxy sample revealed a significant
excess (at the confidence level P=0.988) of gas-rich objects in the
void sample
with respect to similar late-type galaxies residing in the Local Volume (LV)
groups and in the CVnI cloud. For the LV group objects which do not belong to
the CnVI cloud, the difference is significant at the confidence level of
0.9956.
\item
This result can be treated as  independent evidence for the slower
evolution of typical void galaxies. This is consistent with similar
conclusions previously published by the authors, based on the analysis of
gas-phase metallicity in void galaxies and similar galaxies in denser
environments.
\item
The ratio $M$(\HI)/$L_{\rm B}$ for the void galaxies has a broad distribution
with extreme values of $\sim$0.05 and $\sim$28,
indicating that various competing factors can define the galaxy
evolution in voids. The median value of $M$(\HI)/$L_{\rm B}$ varies
with $M_{\rm B}$ within a factor of $\sim$4  (from $\sim$0.5 to $\sim$2) for
a luminosity range of $\sim$3 orders of magnitude.
\end{enumerate}

\begin{acknowledgements}
We thank the NRT Program Committee for allocating time for this project during
the years 2007 -- 2012. The authors appreciate the anonymous referee's
report, which allowed us to improve the quality of the paper.
The work of SAP was partly supported by RFBR grants No.
11-02-00261 and 14-02-00520. SAP appreciates the hospitality and support
from the Paris Observatory and its laboratory GEPI during his visits for this
project. We thank J. Chengalur for permission to use the \HI-fluxes for
UGC~3672 system components prior to publication.
The Nan\c {c}ay Radio Observatory is the Unit\'{e}
scientifique de Nan\c {c}ay of
the Observatoire de Paris, associated as Unit\'{e} de Service et de Recherche
(USR) No.~B704 to the French Centre National de la Recherche (CNRS). The
Nan\c {c}ay
Observatory also gratefully acknowledges the financial support of the Conseil
Regional de la Region Centre in France.
This research has made use of the NASA/IPAC Extragalactic Database (NED)
which is operated by the Jet Propulsion Laboratory, California Institute
of Technology, under contract with the National Aeronautics and Space
Administration. We acknowledge the usage of the HyperLeda database
(http://leda.univ-lyon1.fr, Paturel et al. 2003).

\end{acknowledgements}

\footnotesize{\bf {
\begin{table*}
\caption[]{Parameters of the Lynx-Cancer void sample galaxies observed with NRT}
\label{t:Param}
\begin{tabular}{lllcrrllrl} \hline \hline
\\[-0.3cm]
\multicolumn{1}{c}{Short IAU}& \multicolumn{1}{c}{Other~~~} & \multicolumn{1}{c}{Type~ } &
\multicolumn{2}{c}{Coord. (2000.0)} & \multicolumn{1}{c}{$V_{\rm opt}$} &
\multicolumn{1}{c}{$V$(HI)} &  \multicolumn{1}{c}{$B_{\rm tot}$$\ddagger$} &
\multicolumn{1}{c}{$M_{\rm B}^{0}$$^*$}  &
\multicolumn{1}{l}{Alternative}  \\ \cline{4-5}
 style name & name or &  & R.A. & Dec.~~~ & \kms\   & \kms\  & mag & mag &  name  \\
  & prefix &  &\ \ $^h$\ \ $^m$\ \ $^s$ \ & \ $^{\circ}$~~~\ \ $'$~~~\ \ $''$~~~ &  &    &     &   \\
~~~~(1) & ~~(2) & (3)  & (4)  & (5)~~~~~ & (6)~~ & (7)~~~ & (8) & (9) & ~~~(10)    \\ \hline \\
J0626+2439 &HIPASS & Scd  & 06 26 20.97&  +24 39 20.0 &  1473$\pm$7  &    1485$\pm$2  & 17.98 &--15.64 &                 \\ 
J0629+2334 &HIPASS & Scd  & 06 29 58.23&  +23 34 28.5 &  1452$\pm$6  &    1445$\pm$3  & 17.10 &--15.88 &PGC1689759       \\ 
J0713+2926$^{*}$&SDSS&dI? & 07 13 05.15&  +29 26 42.8 &  ...         &    938$\pm$2   & 16.79 &--14.55 &                 \\ 
J0723+3621 &SDSS   & Sm?  & 07 23 01.42&  +36 21 17.1 &  888$\pm$2   &    917$\pm$2   & 17.01 &--14.21 &                 \\ 
J0723+3622 &SDSS   & LSB  & 07 23 13.46&  +36 22 13.0 &  954$\pm$3   &    970$\pm$5   & 19.46 &--11.76 &                 \\ 
J0730+4109 &SDSS   & dI?  & 07 30 58.90&  +41 09 59.8 &  874$\pm$3   &    878$\pm$5   & 16.67 &--14.59 &                 \\ 
J0737+4724 &SDSS   & LSB  & 07 37 28.47&  +47 24 32.8 &  404$\pm$60  &    474$\pm$5   & 18.02 &--12.54 &                 \\ 
J0744+2508 &SDSS   & dI   & 07 44 43.72&  +25 08 26.6 &  749$\pm$4   &    760$\pm$3   & 18.11 &--12.66 &                 \\ 
J0744+2506$^{*}$&SDSS&dI  & 07 44 55.52&  +25 06 01.8 &  778$\pm$99  &    752$\pm$6   & 20.47 &--10.30 &                 \\ 
J0747+5111 &SDSS   & Sm   & 07 47 32.10&  +51 11 29.0 &  454$\pm$84  &    433$\pm$3   & 15.12 &--15.16 & MCG 9-13-56     \\ %
J0802+0525$^{1)}$ &SDSS&Comp& 08 02 38.15&+05 25 51.2 &  830$\pm$23  &    824$\pm$6   & 19.80 &--10.93 & AGC~188988      \\  %
J0809+4135 &SDSS   & Sd?  & 08 09 36.10&  +41 35 40.0 &  704$\pm$50  &    712$\pm$5   & 15.46 &--15.41 & MCG 7-17-19     \\ 
J0810+1837 &SDSS   & Sm:  & 08 10 30.65&  +18 37 04.1 & 1495$\pm$37  &    1481$\pm$8  & 18.39 &--13.58 &                 \\ 
J0812+4836 &SDSS   & dI   & 08 12 39.53&  +48 36 45.4 &  521$\pm$5   &    522$\pm$4   & 17.36 &--13.08 &                 \\ %
J0825+3201 &SDSS   & Ir   & 08 25 04.90&  +32 01 05.1 &  648$\pm$16  &    647$\pm$6   & 16.91 &--13.73 & KUG 0821+321     \\ %
J0831+4104  &SDSS  & LSB  & 08 31 41.21&  +41 04 53.7 &  582$\pm$40  &    640$\pm$3   & 17.71 &--12.92 &                 \\ 
J0843+4025  &SDSS  & Im   & 08 43 37.98&  +40 25 47.2 &  614$\pm$3   &    627$\pm$10  & 17.90 &--12.68 &                 \\ 
J0845+1519  &SDSS  & dI   & 08 45 25.40&  +15 19 46.0 & 1642$\pm$50  &   1584$\pm$12  & 18.61 &--13.40 &                  \\ 
J0852+1350  &SDSS  & LSB  & 08 52 33.75&  +13 50 28.3 & 1511$\pm$4   &   1502$\pm$8   & 17.40 &--14.56 &                 \\ 
J0859+3923  &SDSS  & dI   & 08 59 46.93&  +39 23 05.6 &  588$\pm$34  &   568$\pm$3    & 17.25 &--13.14 &                 \\ 
J0900+3222$^{*}$&SDSS&dI  & 09 00 18.30&  +32 22 26.2 &  740$\pm$30  &   ...          & 18.97 &--11.77 &                 \\ 
J0911+3135  &SDSS  & dI   & 09 11 59.43&  +31 35 35.9 &  750$\pm$4   &   753$\pm$6    & 18.05 &--12.68 &                 \\ 
J0917+2525  &IC2450& S0   & 09 17 05.27&  +25 25 44.9 & 1644$\pm$2   &   1643$\pm$12  & 14.06 &--18.11 &                 \\ 
J0926+3343  &SDSS  & Sm:  & 09 26 09.45&  +33 43 04.1 &  565$\pm$57  &   536$\pm$2    & 17.30 &--12.91 &                 \\ 
J0928+2845  &SDSS  & dI   & 09 28 59.06&  +28 45 28.5 & 1229$\pm$41  &   1224$\pm$7   & 16.76 &--14.82 &                 \\ 
J0929+1155  &SDSS  & dI   & 09 29 51.83&  +11 55 35.7 & 1349$\pm$51  &   1614$\pm$8   & 17.20 &--14.84 &                 \\ 
J0931+2717  &SDSS  & Sm:  & 09 31 36.15&  +27 17 46.6 & 1505$\pm$2   &   1504$\pm$3   & 18.00 &--13.94 &                 \\ 
J0934+0625$^{*}$&CGCG035-007&Sc&09 34 44.72&+06 25 31.2& 574$\pm$38  &    548$\pm$4   & 15.42 &--14.50 &                 \\ 
J0937+2733  &SDSS  & Im   & 09 37 47.65&  +27 33 57.7 & 1595$\pm$16  &   1588$\pm$1   & 16.50 &--15.58 &                 \\ 
J0940+4459  &SDSS  & dI   & 09 40 03.27&  +44 59 31.7 & 1358$\pm$4   &   1350$\pm$10  & 18.01 &--13.79 &                 \\ 
J0942+09375$^{*}$&SDSS&dI & 09 42 51.25&  +09 37 57.6 & 1461$\pm$17  &   1456$\pm$6   & 18.15 &--13.67 &                 \\ 
J0943+4134  &SDSS  & dI   & 09 43 42.97&  +41 34 08.9 & 1403$\pm$40  &   1436$\pm$4   & 17.64 &--14.25 &                 \\ 
J0944+1000  &SDSS  & dI   & 09 44 37.10&  +10 00 46.3 & 1477$\pm$66  &   1476$\pm$3   & 16.96 &--14.89 &                 \\ 
J0947+4138  &SDSS  & BCG  & 09 47 18.35&  +41 38 16.4 & 1389$\pm$2   &   1400$\pm$2   & 17.92 &--13.93 & HS 0944+4152    \\ 
J0947+3905a$^{*}$&MRK407&BCG&09 47 47.60& +39 05 03.0 & 1589$\pm$10  &   1582$\pm$4   & 15.28 &--16.79 &                 \\ 
J0947+3908$^{*}$&UZC&Sd   & 09 47 50.25&  +39 08 31.7 & 1553$\pm$25  &   1565$\pm$4   & 16.85 &--15.20 &                 \\ 
J0947+3905b &SDSS  & LSB  & 09 47 58.45&  +39 05 10.1 & 1501$\pm$60  &   1567$\pm$4   & 18.03 &--14.02 &                 \\ 
J0951+3842  &SDSS  & dI   & 09 51 41.67&  +38 42 07.3 & 1435$\pm$4   &   1433$\pm$7   & 17.46 &--14.43 &                 \\ 
J0954+3620  &SDSS  & dI   & 09 54 50.60&  +36 20 01.9 &  503$\pm$55  &   550$\pm$5    & 18.05 &--12.17 &                 \\ %
J0956+2716$^{*}$&SDSS&dI  & 09 56 33.65&  +27 16 59.3 & 1074$\pm$25  &   1059$\pm$2   & 18.13 &--13.17 &                 \\ 
J0957+2745$^{*}$&SDSS&dI  & 09 57 29.40&  +27 45 24.3 & 1184$\pm$16  &   1184$\pm$4   & 18.16 &--13.33 &                 \\ %
J0959+4736  &SDSS  & dI   & 09 59 18.60&  +47 36 58.4 & 1093$\pm$4   &   1110$\pm$12  & 17.05 &--14.37 & PC 0956+4751    \\ 
J1000+3032  &SDSS  & dI   & 10 00 36.54&  +30 32 09.8 &  501$\pm$37  &    484$\pm$13  & 18.14 &--11.87 &                 \\ 
J1010+4617  &SDSS  & dI   & 10 10 14.96&  +46 17 44.1 & 1092$\pm$3   &   1092$\pm$3   & 18.23 &--13.15 &                 \\  
J1019+2923  &SDSS  & dI   & 10 19 28.52&  +29 23 02.3 &  874$\pm$43  &    885$\pm$4   & 17.48 &--13.60 &                 \\ \hline 
\\[-0.3cm]
\multicolumn{10}{l}{($^{1)}$) -- probable artifact. See text in Sect.~\ref{sec:results}.  } \\
\end{tabular}
\end{table*}
     }
 }
\normalsize

\footnotesize{\bf {
\begin{table*}
\caption[]{\HI\ parameters of the observed Lynx-Cancer void sample galaxies }
\label{t:HI}
\begin{tabular}{lcrrrrrrrrr} \hline \hline
\\[-0.3cm]
\multicolumn{1}{c}{Short IAU}& \multicolumn{1}{c}{$V$(HI)$\pm \sigma$} & \multicolumn{1}{c}{Dist } &
\multicolumn{1}{c}{$W_{\mathrm 50}\pm \sigma$}& \multicolumn{1}{c}{$W_{\mathrm 20}\pm \sigma$} &
\multicolumn{1}{c}{$F$(HI)$\pm \sigma$ } &
\multicolumn{1}{c}{$\log \pm \sigma$}& \multicolumn{1}{c}{$M$(\HI)$\pm \sigma$} & \multicolumn{1}{c}{Time} &
\multicolumn{1}{c}{RMS } & \multicolumn{1}{c}{S/N} \\ \cline{8-8}
\multicolumn{1}{c}{style name}& \multicolumn{1}{c}{\kms} & \multicolumn{1}{c}{adopt } &
\multicolumn{1}{c}{\kms}& \multicolumn{1}{c}{\kms} & \multicolumn{1}{c}{Jy~\kms } & \multicolumn{1}{c}{$M$(\HI)} &
\multicolumn{1}{c}{$L_{\rm B}$} & \multicolumn{1}{c}{min} & \multicolumn{1}{c}{mJy} & \multicolumn{1}{c}{ }  \\ 
~~~~(1) & ~~(2) & (3)  & (4)  & (5)~~~~~ & (6)~~ & (7)~~~ & (8) & (9) & ~~~(10) & (11)   \\ \hline \\
J0626+2439 &  1485$\pm$2   &23.21 & 138$\pm$3  &156$\pm$5&  7.04$\pm$0.22 &   8.95$\pm$0.013 &   3.20$\pm$0.100   &  42 & 3.5   & 16.2 \\ 
J0629+2324 &  1445$\pm$2   &22.92 & 129$\pm$3  &150$\pm$4& 10.40$\pm$0.25 &   9.11$\pm$0.010 &   3.69$\pm$0.090   &  22 & 4.1   & 22.2 \\ 
J0713+2926 &   938$\pm$2   &16.10 &  40$\pm$3  & 61$\pm$4 & 1.57$\pm$0.06 &   7.98$\pm$0.016 &   0.93$\pm$0.035   &  80 & 1.6   & 24.0 \\ 
J0723+3621$^{*}$&917$\pm$1 &16.00 & 100$\pm$4  &122$\pm$6 & 3.74$\pm$0.18 &   8.35$\pm$0.021 &   2.93$\pm$0.143   &  64 & 3.4   & 17.2 \\ 
J0723+3622$^{*}$&970$\pm$1 &16.00 &  45$\pm$10 & 69$\pm$16& 1.59$\pm$0.16 &   7.98$\pm$0.041 &  11.89$\pm$1.189   &  49 & 3.8   & 7.0  \\ 
J0730+4109 &   878$\pm$5   &15.75 &  51$\pm$10 & 72$\pm$16& 0.74$\pm$0.10 &   7.64$\pm$0.056 &   0.41$\pm$0.056   &  84 & 2.4   & 6.0  \\ 
J0737+4724 &   474$\pm$5   &10.40 &  40$\pm$4  & 53$\pm$6 & 0.99$\pm$0.09 &   7.40$\pm$0.037 &   1.58$\pm$0.140   & 117 & 2.4   & 10.8 \\ 
J0744+2508 &   760$\pm$3   &13.10 &  28$\pm$6  & 43$\pm$9 & 0.39$\pm$0.05 &   7.20$\pm$0.052 &   0.88$\pm$0.113   & 128 & 1.7   & 7.5  \\ 
J0744+2506 &   752$\pm$6   &13.10 &  22$\pm$12 & 34$\pm$19&0.032$\pm$0.016&   6.11$\pm$0.176 &   0.64$\pm$0.320   & 140 & 1.6   & 2.9  \\ 
J0747+5111 &   433$\pm$3   & 9.92 &  75$\pm$5  &105$\pm$9 & 5.96$\pm$0.27 &   8.14$\pm$0.019 &   0.77$\pm$0.035   &  26 & 5.2  & 15.4 \\ 
J0802+0527 &   830$\pm$6   &13.25 & ...        & ...      & 0.10$\pm$0.10 &$<$6.62$\pm$0.301 &$<$1.14$\pm$1.140   &  69 & 2.8   & 2.0  \\ 
J0809+4135 &   712$\pm$5   &13.48 &  95$\pm$10 &137$\pm$16& 5.13$\pm$0.30 &   8.34$\pm$0.025 &   0.97$\pm$0.057   &  19 & 5.1  & 12.0 \\ 
J0810+1837 &  1481$\pm$8   &23.03 &  42$\pm$15 & 64$\pm$24& 0.46$\pm$0.09 &   7.76$\pm$0.079 &   1.37$\pm$0.274   &  25 & 2.3   & 4.4  \\ 
J0812+4836 &   522$\pm$4   &11.04 &  47$\pm$8  & 64$\pm$12& 1.27$\pm$0.18 &   7.56$\pm$0.057 &   1.38$\pm$0.192   & 166 & 4.4   & 6.5  \\ 
J0825+3201 &   647$\pm$6   &12.23 &  62$\pm$12 & 84$\pm$18& 1.11$\pm$0.17 &   7.59$\pm$0.062 &   0.82$\pm$0.126   &  37 & 3.7   & 5.6  \\ 
J0831+4104 &   640$\pm$3   &12.44 &  25$\pm$6  & 32$\pm$9 & 0.17$\pm$0.06 &   6.79$\pm$0.135 &   0.27$\pm$0.098   &  92 & 2.2   & 3.5  \\ 
J0843+4025 &   627$\pm$10  &12.23 &  24$\pm$21 & 46$\pm$33& 0.19$\pm$0.09 &   6.83$\pm$0.168 &   0.37$\pm$0.175   &  63 & 2.8   & 3.2  \\ 
J0845+1519 &  1584$\pm$12  &24.19 &   6$\pm$24 & 39$\pm$38& 0.10$\pm$0.05 &   7.14$\pm$0.176 &   0.39$\pm$0.195   & 175 & 1.6   & 4.1  \\ 
J0852+1350 &  1502$\pm$8   &22.96 &  77$\pm$15 &118$\pm$24& 2.30$\pm$0.22 &   8.46$\pm$0.039 &   2.75$\pm$0.261   &  20 & 4.0   & 8.0  \\ 
J0859+3923 &   568$\pm$3   &11.36 &  31$\pm$7  & 46$\pm$11& 0.60$\pm$0.09 &   7.26$\pm$0.064 &   0.65$\pm$0.103   &  54 & 2.8   & 6.5  \\ 
J0900+3222 &   740$\pm$30  &13.18 & ...        & ...      & 0.14$\pm$0.10 &$<$6.76$\pm$0.231 &$<$0.72$\pm$0.520   &  35 & 3.4   & 1.4  \\ 
J0911+3135 &   753$\pm$6   &13.56 &  27$\pm$12 & 52$\pm$18& 0.48$\pm$0.07 &   7.32$\pm$0.061 &   1.14$\pm$0.171   &  85 & 2.0   & 6.5  \\ 
J0917+2525 &  1643$\pm$12  &25.45 &  98$\pm$24 &131$\pm$38& 0.81$\pm$0.16 &   8.09$\pm$0.079 &   0.05$\pm$0.010   &  23 & 2.8   & 4.1  \\ 
J0926+3343 &   536$\pm$2   &10.63 &  43$\pm$4  & 79$\pm$7 & 2.71$\pm$0.10 &   7.86$\pm$0.015 &   3.23$\pm$0.117   &  72 & 2.2   & 24.4 \\ 
J0928+2845 &  1224$\pm$7   &19.84 &  94$\pm$14 &123$\pm$22& 1.99$\pm$0.26 &   8.27$\pm$0.053 &   1.41$\pm$0.182   &  17 & 4.6   & 6.3  \\ 
J0929+1155 &  1614$\pm$8   &24.29 & 119$\pm$15 &138$\pm$24& 3.14$\pm$0.17 &   8.64$\pm$0.023 &   3.27$\pm$0.178   &  37 & 2.9   & 3.7  \\ 
J0931+2717 &  1504$\pm$3   &23.59 &  48$\pm$6  & 71$\pm$10& 0.54$\pm$0.13 &   7.85$\pm$0.092 &   1.21$\pm$0.285   &  47 & 3.0   & 11.0 \\ 
J0934+0625 &   548$\pm$4   & 8.88 &  86$\pm$8  &116$\pm$12& 4.07$\pm$0.22 &   7.88$\pm$0.023 &   0.77$\pm$0.042   &  34 & 4.1   & 11.4 \\ 
J0937+2733 &  1588$\pm$1   &25.08 &  33$\pm$2  & 52$\pm$3 & 3.04$\pm$0.12 &   8.66$\pm$0.016 &   1.71$\pm$0.065   &  38 & 3.2   & 27.1 \\ 
J0940+4459 &  1350$\pm$10  &22.25 &  30$\pm$16 & 50$\pm$25& 0.11$\pm$0.05 &   7.11$\pm$0.163 &   0.25$\pm$0.114   &  77 & 2.8   & 1.3  \\ 
J0942+0937 &  1456$\pm$6   &21.93 &  43$\pm$13 & 63$\pm$20& 0.55$\pm$0.05 &   7.80$\pm$0.041 &   1.37$\pm$0.137   &  26 & 2.4   & 5.0  \\ 
J0943+4134 &  1436$\pm$4   &23.33 &  46$\pm$9  & 64$\pm$13& 0.43$\pm$0.06 &   7.74$\pm$0.060 &   0.70$\pm$0.104   &  71 & 1.6   & 6.3  \\ 
J0944+1000 &  1476$\pm$3   &22.21 &  64$\pm$6  & 87$\pm$9 & 2.23$\pm$0.14 &   8.41$\pm$0.027 &   1.84$\pm$0.116   &  21 & 3.0   & 12.1 \\ 
J0947+4138 &  1400$\pm$2   &22.71 &  61$\pm$3  & 90$\pm$5 & 0.20$\pm$0.07 &   7.39$\pm$0.132 &   0.42$\pm$0.149   & 132 & 1.5   & 27.9 \\ 
J0947+3905a&  1582$\pm$4   &25.11 & 157$\pm$7  &193$\pm$12& 7.00$\pm$0.50 &   9.02$\pm$0.030 &   1.29$\pm$0.092   &  22 & 3.3   & 27.0 \\ 
J0947+3908 &  1565$\pm$4   &25.11 & ...        & ...      & 5.00$\pm$0.40 &   8.87$\pm$0.033 &   3.92$\pm$0.314   &  20 & 3.0   & 22.0 \\ 
J0947+3905b&  1567$\pm$4   &25.11 & ...        & ...      & 1.50$\pm$0.20 &   8.35$\pm$0.054 &   3.48$\pm$0.464   &  20 & 3.0   & 21.0 \\ 
J0951+3842 &  1433$\pm$7   &23.04 &  29$\pm$14 & 56$\pm$22& 0.48$\pm$0.11 &   7.78$\pm$0.089 &   0.65$\pm$0.148   &  48 & 2.9   & 5.7  \\ 
J0954+3620 &   550$\pm$5   &10.86 &  44$\pm$10 & 63$\pm$15& 0.28$\pm$0.05 &   6.89$\pm$0.074 &   0.68$\pm$0.126   &  86 & 1.3   & 6.0  \\ 
J0956+2716$^{**}$&1059$\pm$2&19.94&  56$\pm$5  & 68$\pm$8&  1.46$\pm$0.12 &   8.14$\pm$0.035 &   1.90$\pm$0.161   &  44 & 3.0   & 10.9 \\ 
J0957+2745 &  1184$\pm$4   &19.14 &  40$\pm$9  & 52$\pm$13& 0.33$\pm$0.07 &   7.46$\pm$0.086 &   0.86$\pm$0.188   &  45 & 2.1   & 4.2  \\ %
J0959+4736 &  1110$\pm$12  &18.89 &  80$\pm$24 &134$\pm$37& 1.91$\pm$0.16 &   8.21$\pm$0.036 &   1.80$\pm$0.154   &  47 & 3.6   & 6.8  \\ 
J1000+3032 &   484$\pm$13  & 9.68 &  34$\pm$25 & 55$\pm$39& 0.15$\pm$0.07 &   6.52$\pm$0.166 &   0.38$\pm$0.177   &  29 & 2.5   & 3.0  \\ 
J1010+4617 &  1092$\pm$3   &18.57 & ...        & ...      & 0.06$\pm$0.06 &$<$6.69$\pm$0.301 &$<$0.34$\pm$0.340   &  37 & 2.4   & 2.0  \\ 
J1019+2923 &   885$\pm$4   &15.40 &  71$\pm$8  & 86$\pm$12& 1.04$\pm$0.13 &   7.77$\pm$0.051 &   1.36$\pm$0.170   &  23 & 2.8   & 5.8  \\ \hline 
\\[-0.3cm]
\multicolumn{11}{l}{($^*$) -- \HI\ parameters are adopted based on GMRT data from Chengalur, Pustilnik (\cite{Chengalur13}); }   \\
\multicolumn{11}{l}{($^{**}$) -- distance as for IC2520=J0956+2713, a massive component of pair with $V_{\rm hel}$=1243~\kms. }   \\
\end{tabular}
\end{table*}
     }
 }
\normalsize


\clearpage

\footnotesize{\bf {
\begin{table*}
\caption[]{Lynx-Cancer void sample galaxies with \HI\ data
from literature}
\label{t:Param_liter}
\begin{tabular}{llllrrrcrrl} \hline
\multicolumn{1}{l}{Short IAU}& \multicolumn{1}{l}{Other~~~} & \multicolumn{1}{c}{Type~ } &
\multicolumn{2}{c}{Coord. (2000.0)} & \multicolumn{1}{c}{$V_{\rm hel}$} &
\multicolumn{1}{c}{$B_{\rm tot}$} &  \multicolumn{1}{c}{$M_{\rm B}^{0}$$^*$} &
\multicolumn{1}{l}{$F$(\HI)}  &  \multicolumn{1}{c}{$M$(\HI)}  & \multicolumn{1}{l}{Source} \\ \cline{4-5}
 style name & name or &  & R.A. & Dec.~~~ & \kms   & mag  & mag & Jy & $L_{\rm B}$ & of \HI   \\
  & prefix &  &\ \ $^h$\ \ $^m$\ \ $^s$ \ & \ $^{\circ}$~~~\ \ $'$~~~\ \ $''$~~~ &  &  &  &\kms  &(sun) &   \\
~~~~(1) & ~~(2) & (3)  & (4)  & (5)~~~~~ & (6)~~ & (7)~~~ & (8) & (9) & ~~~(10) & ~~~(11)   \\ \hline 

J0630+3930 & UGC3475     & Sm  & 06 30 28.86& +39 30 13.6&  487 & 14.97 & -15.88 & 24.60 & 1.76  & HR89   \\ %
J0630+3318 & UGC3476     & Im  & 06 30 29.22& +33 18 07.2&  469 & 14.96 & -16.02 & 12.51 & 0.72  & Sch92  \\ %
J0638+2239 & UGC3503     & Sd  & 06 38 01.40& +22 39 06.0& 1389 & 15.10 & -17.28 &  9.66 & 0.89  & Sp05   \\ %
J0838+4915 & UGC3501     & Im: & 06 38 38.40& +49 15 30.0&  449 & 17.20 & -13.32 &  3.60 & 2.62  & HR89   \\ %
J0643+2252 & UGC3516     & Sd  & 06 43 08.51& +22 52 24.9& 1287 & 16.97 & -15.77 &  2.81 & 0.88  & Sch90  \\ %
J0647+4730 & KKH 38      & I   & 06 47 54.88& +47 30 50.0&  451 & 17.40 & -12.98 &  6.50 & 6.25  & KKH01  \\ %
J0653+1917 & UGC3587     & S?  & 06 53 54.70& +19 17 59.0& 1267 & 13.84 & -18.08 & 49.57 & 1.80  & Sp05   \\ %
J0655+3905 & UGC3600     & Im: & 06 55 40.00& +39 05 42.8&  412 & 15.92 & -14.21 &  5.50 & 1.49  & Sch92  \\ %
J0706+3020$^{*}$ & SDSS  & LSB & 07 06 23.43& +30 20 51.3&  937 & 19.10 & -12.36 &  3.41 &17.10  & CPE16  \\ 
J0706+3019 & UGC3672     & Im  & 07 06 27.56& +30 19 19.4&  994 & 15.92 & -15.54 & 11.94 & 3.15  & CPE16  \\ 
J0709+4422 & UGC3698     & Im  & 07 09 16.8 & +44 22 48.0&  422 & 15.41 & -14.96 &  7.95 & 1.20  & Sw02   \\ %
J0710+4427 & NGC2337     & IBm & 07 10 13.6 & +44 27 25.0&  436 & 13.48 & -16.85 & 38.00 & 1.00  & Sw02   \\ %
J0713+1031$^{*}$&UGC3755 & Im  & 07 13 51.80& +10 31 19.0&  315 & 14.07 & -15.66 & 10.56 & 0.48  & Sp05   \\ %
J0722+4506 & UGC3817     & Im  & 07 22 44.48& +45 06 30.7&  437 & 15.96 & -14.44 & 10.20 & 2.50  & Sw02   \\ %
J0723+3624$^{*}$&SDSS    & LSB & 07 23 20.57& +36 24 40.8&  938 & 21.68 &  -9.56 &  0.48 &28.30  & CP13   \\ 
J0725+0910 & PGC020981   & I   & 07 25 38.95& +09 10 59.8& 1202 & 16.69 & -14.94 &  1.80 & 1.01  & AL11   \\ %
J0727+4826 & UGC3853     & Sdm & 07 27 39.26& +48 26 45.4&  936 & 15.96 & -15.63 &  4.60 & 1.11  & MV00   \\ %
J0728+4046 & UGC3860     & Im  & 07 28 17.2 & +40 46 13.0&  354 & 15.21 & -14.50 & 11.76 & 1.72  & Beg08  \\ %
J0729+2754 & UGC3876     & SAd & 07 29 17.49& +27 54 01.9&  854 & 13.77 & -17.30 & 18.70 & 0.77  & HR89   \\ %
J0734+0432 & UGC3912     & IBm:& 07 34 12.63& +04 32 47.1& 1240 & 14.72 & -16.87 & 14.39 & 1.37  & Sp05   \\ %
J0736+0959$^{*}$&AGC174585& -  & 07 36 10.30& +09 59 11.0&  357 & 17.90 & -11.54 &  0.54 & 1.01  & AL11   \\ %
J0741+4006 & UGC3966     & Im  & 07 41 26.00& +40 06 44.0&  361 & 15.32 & -14.58 & 25.10 & 4.18  & Sw02   \\ %
J0741+1648$^{*}$&DDO47   & IBsm& 07 41 55.00& +16 48 02.0&  272 & 14.89 & -14.78 & 64.00 & 7.73  & Sp05   \\ %
J0742+1633$^{*}$&KK65    & dIrr& 07 42 31.20& +16 33 40.0&  279 & 15.51 & -14.15 &  2.50 & 0.53  & Beg08  \\ %
J0743+0357$^{*}$&CGCG030-012&Sdm:& 07 43 08.77& +03 57 00.0& 932& 15.60 & -15.41 &  3.66 & 0.82  & AL11   \\ %
J0746+5117 & MCG9-13-52  & Sm  & 07 46 56.36& +51 17 42.8&  445 & 16.54 & -13.75 &  2.60 & 1.27  & KKH01  \\ %
J0757+1423$^{*}$&UGC4115 & IAm & 07 57 01.80& +14 23 27.0&  341 & 14.81 & -14.75 & 21.60 & 2.47  & Beg08  \\ %
J0757+3556 & UGC4117     & IBm & 07 57 25.98& +35 56 21.0&  773 & 15.36 & -15.59 &  5.04 & 0.89  & ONe04  \\ %
J0800+4211 & UGC4148     & Sm  & 08 00 23.68& +42 11 37.0&  716 & 15.66 & -15.18 & 12.50 & 2.98  & Sp05   \\ %
J0801+5044 & NGC2500     & SBd & 08 01 53.30& +50 44 15.4&  504 & 12.14 & -18.21 & 34.60 & 0.32  & Sp05   \\ %
J0813+4559 & NGC2537     & BCG & 08 13 14.73& +45 59 26.3&  445 & 12.49 & -17.71 & 21.10 & 0.26  & MU08   \\ 
J0813+4544 & IC2233      & Sd  & 08 13 58.93& +45 44 34.3&  553 & 13.34 & -17.03 & 47.60 & 1.29  & MU08   \\ 
J0814+4903 & NGC2541     & Scd & 08 14 40.18& +49 03 42.1&  548 & 12.34 & -18.36 &135.60 & 1.67  & Sp05   \\ %
J0819+5000 & NGC2552     & SAm & 08 19 20.14& +50 00 25.2&  524 & 13.01 & -17.42 & 28.50 & 0.58  & Sp05   \\ %
J0825+3532 & HS0822+3542 & BCG & 08 25 55.43& +35 32 31.9&  720 & 17.88 & -12.97 &  0.34 & 0.61  & Ch06   \\ %
J0826+3535 & SAO0822+3545& Im  & 08 26 05.59& +35 35 25.7&  740 & 17.74 & -13.11 &  1.00 & 1.58  & Ch06   \\ %
J0826+2145$^{*}$& SDSS   & dI  & 08 26 20.01& +21 45 22.8&  420 & 18.23 & -11.63 &  0.46 & 1.16  & PMM16   \\ %
J0828+4151 & DDO52       & Im  & 08 28 28.53& +41 51 22.8&  397 & 15.35 & -14.87 & 10.80 & 1.96  & Sp05   \\ %
J0859+3912 & UGC4704     & Sdm & 08 59 00.28& +39 12 35.7&  596 & 14.82 & -15.66 & 22.40 & 2.56  & HR89   \\ %
J0900+2536 & UGC4722     & Sdm & 09 00 23.54& +25 36 40.6& 1795 & 15.01 & -17.39 & 11.6  & 1.25  & Ch15   \\ %
J0900+2538$^{*}$&UGC4722C& dI  & 09 00 26.11& +25 38 21.4& 1837 & 17.21 & -15.19 &  4.30 & 4.30  & Ch15   \\ %
J0908+0517 & KKH46       & dI  & 09 08 36.54& +05 17 26.8&  598 & 17.21 & -12.99 &  3.07 & 2.98  & AL11   \\ %
J0929+2502 & SDSS        & dI  & 09 29 00.10& +25 02 57.0& 1661 & 19.03 & -13.16 &  0.36 & 2.38  & AL11   \\ %
J0940+2935 & KISSB23     & Im  & 09 40 12.67& +29 35 29.3&  507 & 16.32 & -13.53 &  2.20 & 1.03  & PM07   \\ %
J0942+3316 & UGC5186     & Im  & 09 42 59.10& +33 16 00.2&  551 & 15.99 & -14.23 &  1.40 & 0.50  & Beg08  \\ %
J0943+3326 & AGC198691   & dI  & 09 43 32.35& +33 26 57.6&  514 & 19.82 & -10.40 &  0.53 & 6.46  & Hir16  \\ %
J0944+0936$^{*}$&IC559   & Sc  & 09 44 43.90& +09 36 55.0&  541 & 14.77 & -15.22 &  5.33 & 0.59  & PB11   \\ %
J0945+3214 & UGC5209     & Im  & 09 45 04.20& +32 14 18.2&  538 & 16.07 & -14.08 &  1.53 & 0.73  & KKH01  \\ %
J0950+3127 & UGC5272b    & Im  & 09 50 19.49& +31 27 22.3&  541 & 17.56 & -12.60 &  1.16 & 1.70  & Sw02   \\ %
J0950+3129 & UGC5272     & Im  & 09 50 22.40& +31 29 16.0&  520 & 14.45 & -15.71 & 19.30 & 1.61  & Sw02   \\ %
J0951+0749$^{*}$&UGC5288 & Sdm?& 09 51 17.20& +07 49 38.0&  556 & 14.62 & -15.61 & 25.30 & 1.96  & PB11   \\ %
J0956+2716$^{*}$&IC2520  & S   & 09 56 20.40& +27 13 39.0& 1243 & 14.27 & -17.32 &  6.04 & 0.43  & HR89    \\ %
J0956+2900$^{*}$& DDO68C & dI  & 09 56 41.07& +29 00 50.7&  506 & 17.48 & -13.12 &  0.73 & 1.00  & Can14  \\ 
J0956+2849 & DDO68A      & Im  & 09 56 45.70& +28 49 35.0&  502 & 14.70 & -15.90 & 26.70 & 2.86  & Ek08   \\ %
J0958+4744$^{*}$&UGC5354 & Sm  & 09 58 53.40& +47 44 13.0& 1168 & 14.15 & -17.39 & 19.91 & 1.30  & Sp05   \\ %
J1004+2921 & UGC5427     & Sdm & 10 04 41.05& +29 21 55.2&  495 & 14.91 & -15.50 &  1.85 & 0.23  & Sch90  \\ 
J1008+2932 & UGC5464     & Sm  & 10 08 07.65& +29 32 30.5& 1011 & 15.77 & -15.47 &  2.86 & 0.81  & Sch90  \\ %
J1016+3746 & UGC5540     & Sc  & 10 16 21.7 & +37 46 48.7& 1166 & 14.63 & -16.85 &  5.37 & 0.54  & PM07   \\ %
J1016+3754 & HS1013+3809 & BCG & 10 16 24.5 & +37 54 46.0& 1173 & 16.02 & -15.46 &  1.51 & 0.86  & PM07   \\ \hline 
\\[-0.3cm]
\multicolumn{11}{l}{$^{*}$ galaxy from the updated sample; {\bf HR89}: Huchtmeier \& Richter 1989; {\bf Sch90}: Schneider et al. 1990; {\bf Sch92}: Schneider et al. 1992} \\
\multicolumn{11}{l}{{\bf Sp05}: Springob et al. 2005;  {\bf KKH01}: Karachentsev et al. 2001; {\bf Sw02}: Swaters et al. 2002; {\bf MV00}: Matthews, van Driel 2000; {\bf CP13}:}  \\
\multicolumn{11}{l}{Chengalur \& Pustilnik 2013; {\bf AL11}:  Haynes et al. 2011; {\bf Beg08}: Begum et al. 2008; {\bf ONe04}: O'Neil 2004; {\bf MU08}: Matthews \& Uson } \\
\multicolumn{11}{l}{2008; {\bf Ch06}: Chengalur et al. 2006; {\bf PMM16}: Pustilnik et al. 2016, submit.; {\bf Ch15}: Chengalur et al. 2015; {\bf Can14}: Cannon et al. 2014; } \\
\multicolumn{11}{l}{{\bf PM07}: Pustilnik \& Martin 2007; {\bf Hir16}: Hirschauer et al. 2016;  {\bf PB11}: Popping \& Braun 2011; {\bf CPE16}: Chengalur et al. 2016.}
\end{tabular}
\end{table*}
     }
 }
\normalsize


\end{document}